\documentstyle[12pt,multicol,amsfonts,epsf,graphicx]{article}

\newcommand\independent{\protect\mathpalette{\protect\independenT}{\perp}}
\def\independenT#1#2{\mathrel{\rlap{$#1#2$}\mkern2mu{#1#2}}}

\newcommand{\cX}{{\cal X}}

\newcommand{\beq}{\begin{equation}}
\newcommand{\eeq}{\end{equation}}
\newcommand{\beqy}{\begin{eqnarray}}
\newcommand{\eeqy}{\end{eqnarray}}
\newtheorem{Observation}{Observation}
\newtheorem{Postulate}{Postulate}
\newtheorem{Definition}{Definition}
\newtheorem{Lemma}{Lemma}
\newtheorem{Theorem}{Theorem}

\newenvironment{Definition*}{{\bf Definition}}{}

\newcommand{\Req}{\stackrel{+}{=}}

\makeatletter
\def\@beginTheorem#1#2{\trivlist \item[\hskip \labelsep{\bf #1\ #2}]}
\def\@opargbegintheorem#1#2#3{ \trivlist
      \item[\hskip \labelsep{\bf #1\ #2\ (#3)}]}
\makeatother

\makeatletter
\def\@beginLemma#1#2{\trivlist \item[\hskip \labelsep{\bf #1\ #2}]}
\def\@opargbeginLemma#1#2#3{ \trivlist
      \item[\hski
 Hence we have the same statements
about the increase of the supports for increasing depth, where
the local transformations are not counted for the depth.
p \labelsep{\bf #1\ #2\ (#3)}]}
\makeatother

\makeatletter
\def\@beginDefinition#1#2{\trivlist \item[\hskip \labelsep{\bf #1\ #2}]}
\def\@opargbeginDefinition#1#2#3{ \trivlist
      \item[\hskip \labelsep{\bf #1\ #2\ (#3)}]}
\makeatother

\makeatletter
\def\@beginCorollary#1#2{\trivlist \item[\hskip \labelsep{\bf #1\ #2}]}
\def\@opargbeginCorollary#1#2#3{ \trivlist
      \item[\hskip \labelsep{\bf #1\ #2\ (#3)}]}
\makeatother

\makeatletter
\def\@beginExample#1#2{\trivlist \item[\hskip \labelsep{\bf #1\ #2}]}
\def\@opargbeginExample#1#2#3{ \trivlist
      \item[\hskip \labelsep{\bf #1\ #2\ (#3)}]}
\makeatother

\def\C{{\mathbb{C}}}
\def\Z{{\mathbb{Z}}}
\def\R{{\mathbb{R}}}

\def\N{{\mathbb{N}}}

\newcommand{\cH}{{\cal H}}
\newcommand{\cG}{{\cal G}}

\title{On causally asymmetric versions \\ of Occam's Razor \\ and their relation to thermodynamics}

\date{July 29, 2008}

\author{Dominik
Janzing\thanks{e-mail:
dominik.janzing@tuebingen.mpg.de}
\\ \small Max Planck Institute for Biological Cybernetics\\
\small  Spemannstr.~38\\
\small  72076 T\"ubingen, Germany 
}

\begin{document}
\maketitle

\begin{abstract}
In real-life statistical data,  it seems that
conditional probabilities for the {\it effect given their causes} tend to be 
less complex and smoother than conditionals for causes, given their effects. 
We have recently proposed and tested methods for 
causal inference in machine learning using a formalization of this principle. 

Here we try to provide some theoretical justification for causal inference methods based upon 
such a ``causally asymmetric'' interpretation of  Occam's Razor. 
To this end, we discuss toy models of cause-effect relations 
from classical and quantum physics as well as computer science in the context of various aspects of complexity.

We argue that this
asymmetry of the statistical dependences between cause and effect has a thermodynamic origin.
The essential link is the tendency of the environment to provide independent background noise
realized by physical systems that are {\it initially} uncorrelated with the system under consideration 
rather than being {\it finally} uncorrelated. This link extends ideas from the literature relating Reichenbach's
principle of the common cause to the second law.
\end{abstract}

\noindent
{\it keywords:} causality, arrow of time, causal inference, non-equilibrium thermodynamics

\section{Causal reasoning from statistical data}

Uncovering non-deterministic causal relations between
observed quantities relies on the evaluation of statistical
dependences and correlations in empirical data. Two types of
statistical data have to be carefully distinguished: in so-called
{\it experimental} data, one observes the change of the distribution
of one variable after interventions that control the value of the
other. More often, one has to evaluate {\it non-experimental} data
where no controlling intervention by the researcher is possible and
he tries to draw causal conclusions merely from observed
dependences in the statistics. Causal reasoning that relies on
non-experimental data is likely to lead to serious misconclusions.
The main obstacle is that statistical dependences between two random
variables $X$ and $Y$ can be due to three types of (non-exclusive) causal relations.
First, $X$ may be a cause of $Y$, second,
$Y$ may be a cause of $X$, or third, there may be a (latent) common
cause, i.e.,  a hidden variable $Z$ effecting $X$ and $Y$.
This is usually referred to as the
``principle of the common cause'' \cite{Reichenbach}. 

\begin{figure}
\epsfysize=2in
\centerline{\epsffile{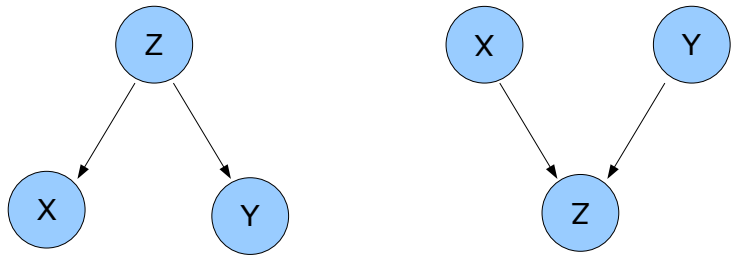}}
\caption{\label{FC} {\small Causal fork (left) and a collider (right), the simplest example of the statistical 
asymmetry between cause and effect (see text)}.}
\end{figure}

If the variables $X$ and $Y$ are time-ordered
and $X$ refers to observations that precedes the observation of $Y$
it is still hard to decide whether $X$ effects $Y$ or there is a
hidden common cause (``confounder'') $Z$. However, it is known that the
joint distribution of at least 3 variables provides some hints on
causal directions \cite{Reichenbach,Spirtes,Pearl:00} via
{\it conditional} independences among variables. For instance,
if  the stochastic dependence between $X$ and $Y$ is only generated
by some common cause $Z$ (see Fig.~\ref{FC}, left),
the variables $X$ and $Y$ must be
independent with respect to the conditional probability given $Z$.
On the other hand, if $Z$ is a common effect of $X$ and $Y$ (see Fig.~\ref{FC}, right), the
role of {\it unconditional} probabilities and {\it conditional}
probabilities is reversed: the conditional probability
given $Z$ would then, in the generic case, generate dependences
between the (actually independent) variables $X$ and $Y$. Common
effects cannot be accepted as an explanation for (unconditional)
dependences but common causes can. Already Reichenbach
\cite{Reichenbach} argued that this statistical asymmetry with
respect to reversing causal arrows is linked to the thermodynamic
arrow of time. Before we describe another asymmetry between cause
and effect that we \cite{SunDJBS} have observed to be useful in causal
reasoning and discuss its relation to statistical physics we first
sketch the known approaches to causal inference from empirical data.

Following \cite{Pearl:00,Spirtes} we restrict our attention to
causal structures without feedback loops and describe
a causal structure as a directed acyclic graph (DAG)
with
random variables
 $X_1,\dots,X_n$ as  nodes.
An arrow from $X_i$ to $X_j$ indicates that $X_i$ directly influences
$X_j$. Even though 
a definition of cause and effect would require deep
philosophical discussions \cite{Pearl:00,Cartwright,Rubin}, we will
define causal relations by referring to {\it hypothetical} interventions.
The variable $X$ influences $Y$ whenever adjusting $X$ (by external
control) to some different value $x$ changes the distribution of
$Y$ (throughout the paper, we will capitalize random variables
and denote their values by lowercase letters). The influence
from $X_i$ on $X_j$ is {\it direct} (relative to the set
$X_1,\dots,X_n$)
 whenever the change of the distribution of $X_j$
caused by different adjustments of $X_i$
 occurs also
 when all the other variables are fixed by an additional intervention.
This definition makes clear
that  causal inference from non-experimental data infers probability
distributions of {\it hypothetical experimental data}.  The connection
between the statistics of non-experimental observations and the causal graph (encoding 
information about the effect of hypothetical interventions)
is provided by the
causal Markov condition. 

\begin{Definition}[causal Markov condition]${}$\\ \label{MKC}
Let $G$ be a DAG with 
$n$ random variables $X_1,\dots,X_n$ as nodes. A joint distribution $P$
on these  variables 
satisfies the (local) Markov condition with respect to $G$ if each variable is, given its parents,
is conditionally independent of its 
non-descendants.
\end{Definition}
Throughout the paper we assume that $P(X_1,\dots,X_n)$ 
has a probability density $p(x_1,\dots,x_n)$ with respect to some product distribution
(note that this assumption does not exclude discrete variables since 
probability mass functions of discrete distributions are also densities).
Then
$p(x_1,\dots,x_n)$ can be factorized into conditional probabilities for each variable, given its parents \cite{Lauritzen}:
\begin{equation}\label{MarkovFac}
p(x_1,\dots,x_n)=\prod_{j=1}^n p(x_j|pa_j)\,,
\end{equation}
where $pa_j$ is a short notation for the subset of values $x_1,\dots,x_n$ that correspond to the parents $PA_j$ 
of $X_j$ with  respectto $G$. The conditional densities $p(x_j|pa_j)$ will  be  called  the
``Markov kernels'' corresponding to the causal hypothesis $G$.
Conversely, every choice of Markov kernels
  $p(x_j|pa_j)$ leads to a Markovian distribution.
 
Following \cite{Pearl:00} we accept  a causal hypothesis only if the observed statistical dependences are consistent
with the Markov condition and mention also that this can be justified by so-called {\it functional models}:

\begin{Definition}[functional  model of causality]${}$\\ \label{functional}
For each node $X_j$ we introduce an
additional noise variable $S_j$ and assume that the actual value $x_j$ of $X_j$ is a deterministic function of
$S_j$ and  all parents of $X_j$. All the $S_j$ are jointly statistically independent.
\end{Definition}

Then the Markov condition follows 
due to  Theorem~1.4.1 in \cite{Pearl:00}. 
It should be noted that noise variable $S_j$ is, by construction, statistically 
independent of the ancestors of $X_j$, but in the generic case there are 
dependences to the descendants of $X_j$.  The relation of this asymmetry between cause and effect to  the  second law
of thermodynamics will be discussed 
in Section~\ref{Sec:CommonRoot}. 

There are at least $n!$ causal graphs for which all
distributions $P$ are Markovian, namely every {\it complete} DAG
(that is, a graph where each $X_i$ has an arrow to  $X_j$ for
every $j>i$ if some arbitrary order of nodes is given). One therefore needs
additional inference rules. So-called independence-based approaches to causal inference
 \cite{Spirtes,Pearl:00} are based upon the
so-called faithfulness assumption:
\begin{Definition}[causal faithfulness condition]${}$\\
A joint probability distribution $P$ on $n$ random variables $X_1,\dots,X_n$
is faithful with respect to $G$ if only those conditional independences are true which are implied by the
Markov condition.  
\end{Definition}
The idea is the following: Given $G$ and an independent choice of values for the free parameters $p(x_j|pa_j)$ it is unlikely to obtain a
non-faithful graph.  It is more natural to assume that an independence relation holds because it is entailed by the causal structure than
that it is due to {\it specific adjustments} of the parameters $p(x_j|pa_j)$. Arguments of this kind are justified by referring to
``Occam's Razor'' \cite{Pearl:00}. Also 
Bayesian methods to causal discovery \cite{HeckermanCausal} are known to 
give an  implicit preference to faithful structures   
provided that the priors are positive densities on the space of all $p(x_j|pa_j)$ \cite{Meek}.

Unfortunately, faithfulness leads rarely to a unique causal
graph. More often there are still several possible causal
hypotheses.
Therefore, additional inference rules are desirable. In seeking new methods
one must be aware of the fact that no inference principle can always lead to correct results
since there is in principle no method to infer causal relations from non-experimental data that is always reliable.
This is because one can construct a technical system with causal structure $G$ 
that generates any desired joint distribution $P$ that factorizes according to eq.~(\ref{MarkovFac}). To this end, let each node $j$ be a random generator
whose inputs are provided by the parents  of $j$  and whose output is sampled according to $p(x_j|pa_j)$. Then the joint output
$x_1,\dots,x_n$ is obviously sampled from $p(x_1,\dots,x_n)$.

Recent proposals  for alternative causal inference methods are based on the observation that in many cases
$p(x_j|pa_j)$ are quite complex functions for one causal directions and simple for others
\cite{Kano,SunDJBS,FriedNach,Comley}.  
Then the idea is  that the causal hypotheses for which the Markov kernels are simpler
are  more likely to be the  true  ones.
We have proposed \cite{ESANNbinary} 
to use this approach for post-selection of causal hypotheses after independence-based algorithms have  already 
reduced the set of potential causal graphs. However, the case of two variables $X$ and $Y$ where the  task is to distinguish between
$X\rightarrow Y$ and $Y\rightarrow X$  is particularly interesting  because  independence-based approaches fail completely. 
We will therefore devote our main attention to this case.

The idea that models in forward (time and causal) direction tend to be simpler than in backward direction, is certainly not new.
The underlying intuition has influenced human 
and automated reasoning since a long time. Psychological studies indicate that human intuition is better in estimating
the strength of causal links (which is encoded in causal conditionals $P(\rm{ effect}|\rm{causes})$) 
than in  inferring non-causal conditionals \cite{Kahneman}. 
For this reason, it is straightforward that simplicity principles (``Occam's Razor'') are automatically interpreted as simplicity of a 
model  when described in  the correct causal direction.
However,
the author is not aware of any systematic exploration of the theoretical background of causally asymmetric interpretations of Occam's Razor from a statistical physics point of view.

The main ideas of this paper can be summarized as follows:

\vspace{0.2cm}
\noindent
(1) The paper describes various simple models from quantum and classical physics where the factorization of the joint distribution 
$P({\rm cause},{\rm effect})$ into $P({\rm cause})P({\rm effect}|{\rm cause})$ yields ``simpler'' terms than
the ``non-causal''  factorization into  $P({\rm effect})P({\rm cause}|{\rm effect})$.
We will  discuss different notions of simplicity for which this is likely to be the case.

\vspace{0.2cm}
\noindent
(2) For these models, we describe why the simplicity of causal and forward-time conditionals  
is because (a) the dynamical laws of motion  and the Hamiltonians are simple, and (b) 
the relevant systems start in statistically independent states rather than ending up in independent states.
This point of  view shows a link  between the suggested asymmetry between cause and effect and the 
arrow of time in thermodynamically irreversible processes. 

\vspace{0.3cm}

While interactions between physical systems $S_1$ and  $S_2$ typically lead to mutual influence,
we can nevertheless obtain well-defined causal directions:

First, the  variable  $X$ will refer to the state of $S_1$ at some time $t$ and $Y$ to the state of $S_2$ 
at some later time $t'>t$ or $X$ and $Y$  refer to different time instants of the same system.

The second approach is to turn the interaction between $S_1$ and  $S_2$ on only after the state of $S_1$ is adjusted to its present state in order to avoid
backaction from $S_2$ to $S_1$.

The third approach
is  to choose physical conditions such that the influence of $S_2$  on $S_1$ is negligible.
We
will, for instance, discuss non-equilibrium steady states with temperature gradient where this is the case.

The paper is organized as follows. In Section \ref{PlMK}
we sketch the inference rule proposed in \cite{SunDJBS}
and our approach to define smoothness of probability distributions
by constrained maximization of conditional entropy. Section~\ref{QModels} describes physical experiments that are consistent
with our inference rule. We discuss how the examples had to be modified if one tries to obtain
simple Markov kernels for  the {\it non-causal} conditionals $P({\rm cause}|{\rm effect})$. 
In Section~\ref{Sec:Aspects} we discuss examples showing  that causal conditionals are also simpler than non-causal ones with respect to other
notions of simplicity, for instance, with respect to computational complexity.
We
consider the computational complexity of
conditional probabilities connecting input and output
of a boolean circuit with additional noise and argue that
 $P({\rm effect}|{\rm cause})$ can efficiently be computed but $P({\rm cause}|{\rm effect})$ cannot,
provided that the inputs are independent. We describe how this
asymmetry is linked to the thermodynamics of computation.

Section~\ref{Sec:CommonRoot} connects the asymmetry between cause and effect in the shapes of conditionals to the asymmetry stated by the
causal Markov condition. 
Section~\ref{NonEq} describes results that link the observed asymmetries to the thermodynamics of non-equilibrium  steady states.

\section{The principle of plausible Markov kernels and its motivation}

\label{PlMK}

To explain our inference principle we consider {\it complete}  DAGs. They are given by an arbitrary ordering
of the $n$ variables and drawing an arrow from each variable to
every other that appears later in the order. Then the causal
hypotheses are uniquely characterized by one out of $n!$ possible
orderings of the nodes (``causal ordering''). 
This is no loss of generality since  the
true  graph can be obtained by removing
statistically irrelevant parents, given that it is a subgraph  of the hypothetical complete graph.  
The {\it Markov
kernels} corresponding to a hypothetical causal order  $X_1,\dots,X_n$ are defined as
the conditional probabilities $p(x_j|x_1,x_2,\dots,x_{j-1})$. 

The venue of our discussion  will be the 
following vague formulation of our inference rule.
\begin{Definition}[plausible Markov kernels method, abstract version]${}$\\ \label{PplMk}
Prefer the hypothetical causal order $X_1,\dots,X_n$ for which the corresponding 
Markov kernels $p(x_j|x_1,\dots,x_{j-1})$ are as simple and smooth as possible.
\end{Definition}
How to
define smoothness and simplicity in  a reasonable way is, however, a difficult
problem. As a first attempt, which provided some encouraging
results, we have chosen the following definition \cite{SunDJBS}.

\begin{Definition}[second order Markov  kernels]${}$\\ \label{SC}
The simplest non-trivial conditionals $p(x_j|x_1,\dots,x_{j-1})$ 
are those that maximize the conditional Shannon entropy
 of $X_j$ given $X_1,\dots,X_{j-1}$ subject
to the given expectations  $E(X_j)=c_j$ and second moments $E(X_jX_i)=d_{ij}$ for $i=1,\dots,j$,
where $c_j$ and $d_{ij}$ denote the ensemble averages of  the corresponding  quantities. 
\end{Definition}

The conditional Shannon entropy  of $X_j$  given $X_1,\dots,X_{j-1}$ is defined  
by $S(X_j|X_1,\dots,X_n):=-\int p(x_1,\dots,x_j)\ln p(x_j|x_1,\dots,x_{j-1}) dx_1,\cdots x_j$, 
where the integral has to be read as a sum for the case of discrete variables.

To include
vector-valued variables $X_j$ with components $X^{(\ell)}_j$ with  $\ell=1,\dots,m_j$, 
one maximizes entropy subject to
the constraints are given by 
\[
E(X^{\ell})_i=c^{(\ell)}_{i} \quad
\hbox{ and } E(X^{(\ell)}_i X^{(k)}_j)=d^{(\ell,k)}_{ij}\,,
\]
for $\ell=1,\dots,m_i$, $k=1,\dots,m_j$.

The term ``second order Markov  kernel'' is justified by the following known fact:

\begin{Theorem}[second order Markov kernels, explicit form]${}$\\
The conditionals given by Definition~\ref{PplMk} 
read
\begin{equation}\label{Expl}
p(x_j|x_1,\dots,x_{n-1})=\frac{1}{z(x_1,\dots,x_{n-1})} \exp\left(\sum_{i\leq j} a_{ij} x_i x_j+bx_j\right)\,,
\end{equation}
with appropriate constants $a_{ij},b$ and the partition function $z(x_1,\dots,x_{j-1})$. 
\end{Theorem}

\noindent
{\it Proof:} We describe the proof for the continuous case, because the discrete one is even more straightforward. 
Let us first assume that the value set of $X_j$ is restricted to a the interval $[-\lambda,\lambda]$.
Then we can define a uniform distribution $U(X_j|X_1,\dots,X_{j-1})$ with density $u(x_j|x_1,\dots,x_{j-1}):=1/(2\lambda)$. 
Maximizing Shannon entropy is then equivalent to minimizing the Kullback-Leibler distance
\begin{eqnarray*}
&&D(P(X_j|X_1,\dots,X_{j-1})\|U(X_j|X_1,\dots,X_{j-1})):=
\\&&\int p(x_1,\dots,x_j)\ln p(x_j|x_1,\dots,x_{j-1})/u(x_j|x_1,\dots,x_{j-1}) 
dx_1\cdots dx_j\,,
\end{eqnarray*}
subject to the same constraints.
Using Theorem~2.2 in \cite{FrieGupt:2006}, the solution is given eq.~(\ref{Expl}). With $\lambda \to \infty$ we obtain the
same solution without restricting $X_j$ to a compact interval.
$\Box$. 

\vspace{0.3cm}
Up to the partition function, the conditionals in eq.~(\ref{Expl}) are given by second order  polynomials.
Since first order polynomials cannot describe statistical dependences between  variables, we have indeed the simplest
non-trivial class of conditionals in the hierarchy \cite{Amari2001} when we define models of $k$th order
as those containing polynomials up to
degree $k$.

\vspace{0.3cm}
Our inference rule reads:
\begin{Definition}[causal inference via second order Markov kernels]${}$\\ \label{SOexpl}
Estimate the first and second moments $E(X_i)$ and $E(X_iX_j)$ from the data set  using the ensemble averages.
For all hypothetical causal orders $X_1,\dots,X_n$ compute the second order Markov kernels
$p(x_j|x_1,\dots x_{j-1})$
in the sense of Definition~\ref{SC} by maximizing conditional entropies subject to these moments. Decide 
by appropriate statistical tests for which ordering the 
obtained joint density $p(x_1,\dots,x_n)$  provides the best fit to the observed data. 
\end{Definition}

This approach
should only be considered as  a preliminary attempt to formalize simplicity.  
Instead of only describing {\it the simplest} conditionals as  above we have also
proposed 
\cite{NeuroComp} a method to quantify the  complexity of conditional densities. Then  
causal inference is done by
preferring the 
direction that minimizes the sum of the complexities of all Markov  kernels.
However, we will focus on  the first approach.

We describe two instances where
our principle is very intuitive.
First we
consider two random variables $X$ and $Y$
where $X$ is binary, i.e., its value set is $\{0,1\}$
and the value  set of $Y$ is $\R$.
Assume that $X$ influences $Y$.
The best second order model for $p(x)$ is just the distribution given by the observed
relative frequencies.
Using eq.~(\ref{Expl}) we obtain
\begin{equation}\label{GaussPl}
p(y|x)=\frac{1}{\sqrt{2\pi a^{-1}}} e^{a y^2+bxy +c}   
\end{equation}
with appropriate $a,b,c$.
For every $x$, $p(y|x)$ is  a  Gaussian distribution where $x$ determines the means \cite{SunDJBS}.
Indeed, after having observed that the
marginal distribution of $Y$ is
a mixture of two Gaussians and that the conditionals $P(y|x)$ are simple
Gaussians
it seems very plausible
to assume that
$X$ effects $Y$ and not vice versa. 

Now we consider the reverse situation where  $Y$
influences  $X$.
The second order  model for $p(y)$ generates the Gaussian distribution.
For $p(x|y)$ we obtain
\cite{SunDJBS}
\begin{equation}\label{tanh}
p(x=1|y)=\frac{1}{2}\Big(1+\tanh(a y+b)\Big)
\end{equation}
with appropriate $a,b \in \R$.
For this example, one checks easily that only the trivial case $X\independent Y$
can have a second order model in both directions.

In Section~\ref{QModels} we will  analyze examples from physics with one binary and one continuous variable that are  consistent with the above second order Markov kernels. 
We have decided to choose examples from quantum mechanics for two reasons. First, the quantum world provides us with natural realizations of 
binary variables. Second, the simplicity of the models under consideration is intriguing.
Nevertheless, quantum {\it superpositions} are not relevant for the arguments in the next subsections.

\section{Second order Markov kernels in physical models}
\label{QModels}

\subsection{Stern-Gerlach experiment}

\begin{figure}\label{SternGerlachFig}
\epsfysize=3in
\centerline{\epsffile{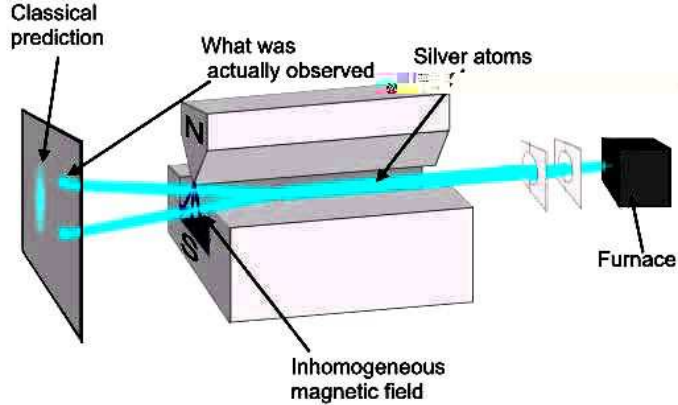} }
\caption{{\small Stern-Gerlach experiment. 
The atom beam emitted by
the furnace splits up into two beams according to the spin.}}
\end{figure}

Consider first an experiment like the one
designed by Stern and Gerlach in 1922  \cite{Stern} to prove the
quantization of  the  magnetic moment.
A beam of atoms is emitted from a furnace and
enters an inhomogeneous magnetic
field  perpendicular to the beam (see Fig.~\ref{SternGerlachFig}, here the field is in vertical direction\footnote{Diagram drawn by  Theresa Knott, taken from the free encyclopedia {\it wikipedia}}),
The field induces a force in the direction of its gradient
which is proportional to the magnetic moment of the particles.
For spin-1/2 particles, for instance, the magnetic moment can attain the values
$+1/2,-1/2$ causing forces in opposite vertical directions.
This effect
can be used as a measurement apparatus for the quantum observable {\it magnetic moment} since it separates the beam
into two parts that hit the screen at different vertical positions. 
We consider the
values $\pm 1/2$ as the two values of a binary variable $X$. Even though
quantum mechanical observables are in general not
random variables on a
probability space, this is well-justified because the quantum superposition  already becomes incoherent 
(by creating entanglement with the position degree of freedom)  when the beam {\it begins} to split up. 
We define furthermore a random
variable $Y$ for the vertical
coordinate of the point where the atom hits the screen.
It is natural to assume that the conditional probabilities
$
p(y|x=\pm 1/2)
$
are both not too different from normal distributions.
The following extremely simplified model, for instance, yields Gaussian conditionals. 
Before the particles have left the source they are
subjected to some
focusing forces. For simplicity
we restrict our attention to the focus in vertical
direction and assume that the forces are induced by a harmonic potential
in vertical direction.
In thermal equilibrium, the
probability distribution of momenta
in a classical as well as in a quantum harmonic oscillator
is Gaussian (see Section~\ref{Thermo} and \cite{WM}, respectively). Assuming that the probability distribution of
the particle momenta is still Gaussian when they leave the source
we obtain for both spin values
Gauss distributions for $Y$ with different expected values.

Even though the terms {\it cause} and {\it effect} are even more philosophically problematic
when quantum effects come into play,
we claim  that $X$ influences $Y$: 
if  we subject a spin measurement to the particles before the beam passes the inhomogeneous field
and remove all atoms with spin down, for instance, we get only one branch of the beam. 
Given the simplified assumptions above, the Markov kernel $p(y|x)$ 
coincides with the second order kernel in eq.~(\ref{GaussPl}). 

Assume now, we had observed a Gaussian
{\it  marginal} distribution for $Y$ instead of Gaussian {\it conditionals}
and a conditional $p(x|y)$ as in eq.~(\ref{tanh}).
Our inference rule would then assume that $Y$  is the cause.
For this reason we want to check whether there are modifications
of the Stern-Gerlach experiment which keep the causal direction but
generate such a distribution.
We could, for instance, assume that the transversal potential
in the furnace is strongly anharmonic
such that the particle momenta $Z$ 
are distributed according to some probability
density
$q(z)$ after the particles have left the source.
Due to the laws of motion, we assume that
$y$ is a linear function in $z$  for both  spin values:
\[
y= a z +  b_+ \hspace{1cm}\hbox{ and } \hspace{1cm} y=az +b_-\,.
\]
Here $b_\pm$ denotes the shift of the expected values
caused by the magnetic moments of particles with spin
$x=\pm 1/2$ and $a\in \R$ is some constant.
In order to obtain Gaussian marginals for $Y$, $q(z)$
must be such that the convex sum of $q(z)$  and its shifted copy is Gaussian.
To see that this is impossible we recall that the Gaussian measure could
then be written as a convolution of $q(z)$ with a measure $\mu$
that is supported by two points. Hence the Fourier transform of $\mu$ multiplied with the
Fourier transform of $q(z)$ would  be the Fourier transform of a Gaussian
which is again a Gaussian (up to a phase function). But this is in contradiction to the fact that
the Fourier transform of $\mu$ has zeros.
This shows that Gaussian marginals for $Y$ cannot be obtained by choosing 
a ``contrived'' potential only. We would also have to modify the laws of motion
given by the magnetic field. One could, for instance,
have a field with strongly inhomogeneous field gradient such
that atoms with different transversal momenta
enter locations with different field gradient.

\begin{figure}
\epsfysize=2in
\centerline{\epsffile{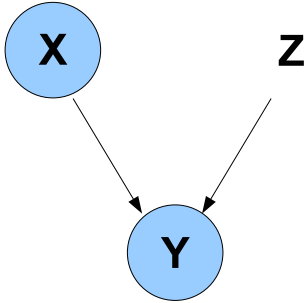}}
\caption{\label{SGG} {\small  Graphical model of the causal structure of the Stern-Gerlach experiment (see text), where the initial momentum represents the noise (that is not explicitly considered in the  causal structure $X\rightarrow Y$).}}
\end{figure}

Fig.~\ref{SGG} shows a simplified graphical model of the causal structure:
the position $Y$ is here assumed to be a deterministic function $f(x,z)$
of the binary variable spin $X$  and a 
``noise'' variable $Z$, the initial momentum. 
Smoothness of the conditional $p(y|x)$ is here due to the smoothness of $f$ and the smoothness of the
distribution of momenta. Last but not least, we should stress the decisive assumption that spin and {\it initial} momenta
are statistically independent.

\subsection{Spin in a stationary magnetic field}

Now we present an example where a continuous classical variable $Y$ influences
a discrete variable $X$. Given a spin-1/2 particle subjected to a field in
z-direction whose (randomly fluctuating) strength
is represented by the
random variable $Y$. The binary variable $X$ represents here the 
possible outcomes $X=\pm 1/2$  for a spin measurement in $z$ direction. 
They occur
in thermal equilibrium 
with the Boltzmann  probabilities, i.e., we have 
\begin{equation}\label{Boltzmann}
p(x=1/2\,|\,y)=\frac{\exp(-ay)}{\exp(-ay)+1}=\frac{1}{2}
\Big(1+\tanh \frac{-a
y}{2}\Big)\,,
\end{equation}
with an appropriate constant $a$ containing Boltzmann's constant $k$, temperature
and the magnetic moment. This is because the density operator of a quantum system with
Hamiltonian $H$ and temperature $T$ is given by
\begin{equation}\label{Gibbs}
\rho=\frac{1}{z} e^{-\frac{1}{kT} H} \,,
\end{equation}  
where $z$  is the appropriate normalization factor.
The conditional probability for the effect
$X$ given the cause $Y$ then is the second order model
in eq.~(\ref{tanh})).

For a $d/2$-spin system having the $d+1$ possible  values
$j=-d/2,-d/2+1,\dots,d/2$ for the spin in a given direction,
second order models 
are conditionals of the form in eq.~(\ref{GaussPl}), i.e.,
\[
p(x=j\,|\,y)=\frac{\exp(-ajy +bj^2+c)}{\sum_{l=-d/2}^{d/2}
\exp(-aly +bl^2+c)}\,,
\]
(with appropriate constants $a,b$) as ``plausible'' conditional distributions.
This parametric family contains the
physically correct
Boltzmann probabilities
\[
p(x=j\,|\,y)=\frac{\exp(- ajy)}{\sum_{l=-d/2}^{d/2}
\exp(- alx)}\,,
\]
by setting $b=c=0$.

As in the Stern-Gerlach experiment, we try to modify the setup (for spin 1/2)   
such that the same causal mechanism leads to a second order model in the
{\it opposite} causal direction.
Then 
$p(y)$  would be a mixture of two Gaussians with equal variance $\sigma^2$
but different expected values $m_\pm$, i.e.,
\begin{eqnarray*}
p(y)=\frac{1}{\sqrt{2\pi} \sigma} &\times& \Big[
p( x=1/2)\, \exp\Big(-\frac{(y-m_+)^2}{2\sigma^2}\Big) \\
&+&p( x=-1/2)\, \exp\Big(-\frac{(y-m_-)^2}{2\sigma^2}\Big)
\Big]\,,
\end{eqnarray*}
To have a field strength that is a mixture of two Gaussians is a priori not unphysical even though
it occurs probably less often than having a unimodal field strength. 
In order to generate the corresponding Gaussian conditionals $p(y|x=\pm 1/2)$ we
had to choose the constant $a$ in eq. (\ref{Boltzmann}) such
that
\[
\frac{p(x=1/2\,|\,y)}{p(x=-1/2\,|\,y)}
=\frac{p(x=1/2)}{p(x=-1/2)} \,\exp\Big(\frac{-(y-m_+)^2 -(y-m_-)^2}{2\sigma^2}\Big)\,.
\]
Comparing this to the Boltzmann probabilities
in eq.~(\ref{Boltzmann}) we conclude that the temperature has to be
chosen such that the constant $a$ satisfies $a/2=2(m_- -m_+)$.
In contrast to the modifications
in the Stern-Gerlach experiment that were required to ``outsmart our principle''
the causal mechanism  as such has not to be modified here.
There is
nevertheless a constraint that makes the described
situation unlikely to occur unless the setup was designed {\it by hand}: The fact that the temperature
value has to be adjusted to {\it one specific} value that is derived from
$m_\pm$  (even though there is no physical reason that makes this coincidence likely)
shows that the counterexample is  {\it non-generic}\footnote{In \cite{Algorithmic} we have argued that this implies that
$P({\rm cause})$ and $P({\rm effect}|{\rm cause})$ share algorithmic information which suggests to prefer the opposite causal  direction.}.

Here,
the reason why thermodynamics predicts a smooth conditional probability for the effect given the cause is, abstractly speaking,
the following. The equilibrium  states maximize entropy subject to the energy.  Here the energy depends smoothly (just linearly) on
the cause (i.e. the field strength). Hence the smoothness of the conditionals is due to the smoothness of
the physical Hamiltonian.

\subsection{Thermal equilibrium with artificial adjustments}

\label{Thermo}

The following setup may be a bit artificial from the physics point of view.
However, it provides a first impression
on the link between the causal direction and the order of maximizing the entropies of subsystems
that is essential for our first implementation of the plausible  Markov kernel principle.
Given two classical systems described by continuous variables $X,Y$
and a joint Hamiltonian of the form
\[
H(x,y)=H_1(x)+H_2(y)+H_{12} (x,y)\,.
\]
Consider the following three hypothetical experiments. For  reasons of convenience, we will identify the systems
with the variables representing the physical states.

\subsection*{(1) System  X and Y influence each other}

Subject the joint
system to a thermal bath with inverse temperature $\beta$.
If  it is thermalized, its statistical state is given by
\[
p_\leftrightarrow (x,y):=\frac{1}{z}\exp\Big(-\beta (H_1(x)+H_2(y)+H_{12} (x,y))\Big)\,,
\]
where $z$ is the partition sum.

\subsection*{(2) System X influences Y}

Remove the interaction term $H_{12}$, subject system $1$ to the bath,
adjust the state
of system $1$, i.e., fix the actual
value  $x$ of the variable $X$. Couple  both systems
by the interaction $H_{12}$ and subject the joint system (or only system 2) to the bath.
Thermalization leads to
\[
p_\rightarrow (x,y)=p_\rightarrow (x)p_\rightarrow (y|x)\,,
\]
with
\[
p_\rightarrow (x):=\frac{1}{z_1}\exp\Big(-\beta H_1(x)\Big)\,,
\]
where $z_1$ is the corresponding partition integral
and
\[
p_\rightarrow (y|x):=\frac{1}{z_\rightarrow (x)}\exp\Big(-\beta (H_2(y)+H_{12} (x,y)\Big)\,,
\]
with the
partition function
\begin{equation}\label{zright}
z_\rightarrow (x):=\int\exp\Big(-\beta (H_2(y)+H_{12} (x,y)\Big) dy\,.
\end{equation}

\subsection*{(3) System Y influences X}
Let $p_{\leftarrow}(x,y)$ be  the density generated by
the same scenario (2) with interchanging the roles of system $1$ and $2$.

\vspace{0.7cm}

Experiment 1 describes bidirectional influence, in 
experiment 2 $X$ is the cause and $Y$ the effect and in experiment 3 we have the reversed case.
Now we want to discuss 
under which circumstances the three  distributions coincide.
For simplicity, we denote by $\Req$ equality up  to an additive constant for the logarithm of unconditional
distributions. We  have certainly $q(x)=r(x)$ if and only if  $\ln q(x)\Req \ln r(x)$.
In analogy to eq.~(\ref{zright}), we introduce the partition function
\[
z_\leftarrow (y):=\int e^{-\beta (H_{12}(x,y)+H_1(x))} dx\,.
\]
We then have
\begin{eqnarray}\label{partDright}
\ln p_\rightarrow (x,y)&\Req&-\beta  \Big(H_1(x)+H_{12}(x,y)+H_2(y)+z_\rightarrow (x)\Big)\\
\ln  p_\leftarrow (x,y)&\Req & -\beta \Big(H_1(x)+H_{12}(x,y)+H_2(y)+z_\leftarrow (y)\Big) \label{partDleft} \\
\ln p_\leftrightarrow (x,y)&\Req & -\beta \Big(H_1(x)+H_{12}(x,y)+H_2(y)\Big)\,. \label{partDboth}
\end{eqnarray}

One checks easily
that $p_\leftrightarrow (y|x)=p_\rightarrow (y|x)$ and $p_\leftrightarrow(x|y)=p_\leftarrow(x|y)$.
Hence the difference between $p_\leftrightarrow $ and $p_\rightarrow$ is only caused by different 
marginal distributions for $X$. While $p_\rightarrow (x)$ is directly determined by the free Hamiltonian $H_1(x)$,
the computation of $p_\leftrightarrow (x)$ involves the partition function $z_\rightarrow (x)$.
We obtain:
\begin{equation}\label{Diff}
\ln p_\leftrightarrow (x,y) -\ln p_\rightarrow (x,y)= -\ln z +  \ln z_1 +\ln z_\rightarrow (x)=:f(x)\,,
\end{equation}
We see that here the partition functions are ``responsible'' for
the fact that different causal directions lead to
different joint distributions because the logarithms  of probabilities are
Hamiltonians  up to a complex function of the  cause.
The following theorem shows under which circumstances the different scenarios yield different joint distributions:

\begin{Theorem}[asymmetries caused by partition function]${}$\\
The following conditions are necessary and sufficient that the joint distributions in the above scenario coincide: 

\begin{enumerate}
\item
$p_\rightarrow=p_\leftrightarrow $ if and only if the partition function $z_\rightarrow$ is a constant
and $p_\leftarrow=p_\leftrightarrow$ if and only if $z_\leftarrow$ is constant.

\item 
$p_\rightarrow=p_\leftarrow$ if and only if both $z_\rightarrow$ and $z_\leftarrow$ are constants.

\item the equalities
$p_\rightarrow (y|x)=p_\leftrightarrow(y|x)$ and $p_\leftarrow(x|y)=p_\leftrightarrow(x|y)$ always hold.

\end{enumerate}
\end{Theorem}

\vspace{0.3cm}
\noindent
{\it  Proof:}
The first part of  the first statement follows by combining eqs.~(\ref{partDright}) with (\ref{partDboth}), the  
second part follows  from symmetry arguments.
The second statement follows from combining eqs.~(\ref{partDright}), (\ref{partDleft}), and (\ref{partDboth}) 
and the fact that a function depending on $x$ can only be
equal to a function of  $y$ up  to a  constant if both functions are constants. 
$\Box$

\vspace{0.3cm}
The above scenarios 
are an example where the joint distribution is obtained by first maximizing the entropy of the cause variable subject to 
the corresponding free Hamiltionian and then maximizing conditional entropy of the effect, given the cause, subject to the  
joint Hamiltonian. In other words, the order of maximizing the entropies coincide with the causal order. 

To see that natural Hamiltonians often will lead to second order Markov kernels,  
let
system $1$ and $2$  be systems with many degrees of freedom, i.e., $x$ and  $y$ are vector-valued variables $(x_1,\dots,x_n)$ and $(y_1,\dots,y_m)$. 
The set of Hamiltonians that occur, are often quadratic terms in the relevant variables (e.g. the canonical variables
$q_i,p_i$). Then  the free Hamiltonians are of the form 
\[
H_1(x)=\sum_j \alpha_j x_j+\sum_{ij} \theta_{ij} x_ix_j\,,
\]
and similarly for $H_2(y)$. 
An important class of possible interaction is given by 
\[
H_{12}(x,y)=\sum_{i,j=1}^n \gamma_{ij} x_ix_j  +\sum_{i,j=1}^m\eta_{ij} y_i y_j +\sum_{i=1}^n\sum_{j=1}^m
\epsilon_{ij} x_i y_j\,,
\]
with parameters $\gamma_{ij},\eta_{ij}, \epsilon_{ij}$.
A natural example would be $d+f$  linearly coupled harmonic oscillators where $n=2d$ and $m=2f$ and 
the $x_i$ are positions and  momenta for $d$ oscillators and $y_i$ for the remaining  $f$ oscillators.

For anharmonic oscillators, one could also have polynomials of higher degree. 
To discuss an example of a system where the Hamiltonian is not a polynomial in the canonical  variables, we  recall
that the potential  energy of an electron in a coloumb field of a positive particle is proportional to $1/x$ where  $x$ is the distance to the particle.
The total energy thus is thus a sum of a polynomial of second order (the kinetic energy) and the $1/x$ term. 

However, second order polynomials already provide a  class  of systems that occur quite often. 
The following theorem is a simple conclusion from the above  remarks.
Its intention is
to stress that the simplicity of Hamiltonians is inherited to the {\it causal} conditionals $P({\rm effect}|{\rm cause})$
but not necessarily to the non-causal ones.

\begin{Theorem}[secord order Markov kernels in equilibrium]${}$\\
Let $S_1$ and $S_2$
be two classical physical  systems with observables $x:=(x_1,\dots,x_n)$ and $y:=(y_1,\dots,y_m)$ 
and assume that their free Hamiltonians  $H_1(x)$, $H_2(y)$ and their interaction Hamiltonian   
$H_{12}(x,y)$ are polynomials of second order. Let system $S_1$ causally influence system $S_2$
in the sense of the above scenario where the state  of $S_1$ is adjusted to the observed value before 
the interaction with $S_2$ is turned on.  Then $p(x)$ and $p(y|x)$ are second order Markov kernels.
\end{Theorem}

\subsection{Stationary process with temperature gradient}

\label{Arm}

In the preceding section, the well-defined causal arrow was put in by hand.
Now we  discuss a natural physical scenario
where back  action is negligible and show that the order of entropy maximization also
coincides with the causal order\footnote{This subsection is related to \cite{MitArmen}, where 
we have considered a model with two interacting systems with non-equilibrium states.
After we assumed separation of dynamical time-scales, 
a well-defined causal arrow emerged whenever the interaction is sufficiently weak compared to the free Hamiltonian of the system  that acts as a cause. In this limiting case, the stationary of the joint system has the following properties.
The state of the system representing the {\it cause} 
was given by a microcanonical distribution of its free Hamiltonian and the state of the system representing the {\it effect} by a microcanonical distribution of its conditional Hamiltonian. Apart from this, it turned out that in the described limit the thermodynamics 
of the ``cause-system'' is a well-behaved thermodynamic system whose coarse-grained entropy is only increasing but never decreasing. 
Hence the conditions to have well-defined thermodynamic properties of subsystems turned out to be related to having well-defined causal directions. However, the setting discussed in the present paper is more appropriate to motivate the method in 
Definition~\ref{SOexpl}}.
To this end, we present a 
model consisting of two baths with different
temperatures. 

Following \cite{Armen2T}
we consider two classical systems 1 and 2, described by  variables
$X$ and $Y$, respectively, and a Hamiltonian $H(x,y)$.
System $j$  is subjected to temperature $T_j$.
Then \cite{Armen2T} describes the coupled Langevin equations
\begin{eqnarray}\label{LangX}
\Gamma_1 \dot{x} &=& \partial_x H(x,y) +\eta_1(t)\\
\Gamma_2 \dot{y} &=& \partial_y H(x,y) +\eta_2(t) \label{LangY}
\end{eqnarray}
with stochastic forces $\eta_j$ whose product satisfies 
\[
E(\eta_i (t) \eta_j (t))  = 2\Gamma_i T_i \delta_{ij}
\delta (t-t')\,
,
\]
where $\Gamma_j$ is the damping constant for system $j$ and $\partial_x$ and $\partial_y$ denote partial derivatives.

Now  $x$
is assumed to change more slowly than $y$ which is ensured by the
condition $\Gamma_1\ll \Gamma_2$. Then it is argued that one may
keep $x$ fixed and solve equation~(\ref{LangY}) for $Y$ and obtain the
$x$-dependent equilibrium
\begin{equation}\label{Pygx}
p(y|x)=\frac{1}{z(x)} \exp\Big(-\beta_1 H(x,y)\Big)\,,
\end{equation}
with the partition function
\[
z(x):=\int \exp\Big(-\beta_1 H(x,y)\Big) \,dy\,,
\]
 and  the inverse temperatures  $\beta_j:=1/(kT_j)$.
In order to calculate $p(x)$  we
average the energy value $H(x,y)$ according to  
$p(y| x)$ in eq.~(\ref{Pygx}) and obtain from eq.~(\ref{LangX}) the Langevin equation
\[
\Gamma_1 \dot{x} = \delta_x H_{{\rm  eff}}(x) +\eta_1(t)
\]
with the effective Hamiltonian
\[
H_{{\rm eff}}(x):= -kT_2\ln \int \exp(-\beta_2 H(x,y))\, dy \,.
\]
Then we obtain 
\begin{equation}\label{MargEff}
p(x) =\frac{1}{z} e^{-\beta_1 H_{{\rm eff}}(x)} 
\end{equation}
and compute the joint distribution using
\[
p(x,y):=p(x)p(y|x)\,.
\]
As has been shown in \cite{Armen2T} that $p(x,y)$ can be obtained by maximizing $T_1 S(X)+ T_2 S(Y|X)$ subject to
\[
\int p(x,y) H(x,y) dxdy =e\,,
\]
for an appropriate value $e$.
This indicates that the limit $T_1\gg T_2$ yields a joint distribution that is obtained by {\it first} maximizing the entropy of system 1 and {\it then} maximizing
the conditional entropy of system 2. 
To study this limit we write
\[
H(x,y)=H_1(x)+H_2(y)+H_{12}(x,y)\,.
\]
Obviously, we have
\begin{equation}\label{Heff}
H_{{\rm eff}}(x)=H_1(x)-kT_2 \int e^{-\beta_2 (H_{12}(x,y)+H_2(y))} dy\,.
\end{equation}

Now  we consider the regime where 
the interaction
$kT_1$ but not small
compared to $kT_2$. Then, intuitively speaking, system 1 does not feel the interaction $H_{12}$, but influences system 2 via $H_{12}$. 
Formally, we consider a sequence of temperatures $T^{(n)}_1:=nT_1$ 
and rescale the free Hamiltonian of system 1 by defining $H_1^{(n)}(x):=nH_1(x)$. The interaction energy and $T_2$ will be kept constant. 
With $\beta_1^{(n)}:=\beta_1/n$ and eq.~(\ref{Heff}) we obtain
\begin{equation}
H^{(n)}_{{\rm eff}}(x)=nH_1(x)-kT_2 \int e^{-\beta_2 (H_{12}(x,y)+H_2(y))} dy\,,
\end{equation}
which yields
\[
e^{-\beta_1^{(n)} H_{{\rm eff}}^{(n)}(x)} \rightarrow e^{-\beta_1 H_1(x)}\,.
\]
Using eq.~(\ref{MargEff}),
the sequence of marginal distributions converge to
\[
p(x)=\frac{1}{z'}e^{-\beta_1 H_1(x)}\,.
\]
We conclude: If $kT_1$ is large compared to the interaction energy the joint distribution 
of the bipartite system is obtained by (1) maximizing the  entropy of system 1 subject to the energy corresponding to its {\it free} Hamiltonian
and then maximizing the conditional entropy of system 2 subject to the total energy. 
We obtain the same statement by decreasing $H_{12}$, $H_2$ and $T_2$ according to a common scaling factor. 

The fact that in these limits the distribution of $X$ is determined by the free Hamiltonian alone is, from an  intuitive perspective, already 
 a good indicator for the fact that the influence of system 2 on system 1 goes to zero.
But our intention is to {\it support} this way of reasoning, not to takes it for granted. 
In order to show that 
 we may indeed consider the variable $X$ as the cause and variable $Y$ as the effect (in the above limits), we show that
system 1 is insensitive with respect to adjusting system 2 to different values as in the preceding subsection.  
To quantify the influence of $Y$ on $X$ we derive an upper bound on the relative entropy distance between
the following two distributions
(1)
the distribution $p_0(x)$ that would be obtained for system 1 without interaction
and (2) the distribution $p_\leftarrow (x|y)$ that system 2 induces when it is adjusted to some
specific value $y$:
\begin{Lemma}[upper bound on the back action]${}$\\ \label{BackBound}
Let $p_\leftarrow$ be defined as in Section~\ref{Thermo} and $y$ be arbitrary, but fixed. 
Define 
\[
p_0(x) =\frac{1}{z} \exp(-\beta_1 H_1(x))\,.
\]
If the interaction energy $H_{12}(x,y)\geq 0$ for all $x$  we have
\[
D\left(P_0(X)||P_\leftarrow(X|y)\right) \leq  \int p_0(x) \beta_1 H_{12}(x,y) \,dx \,,
\]
\end{Lemma}
This shows that the back action indeed converges to zero for $\beta_1 \to 0$. 

\vspace{0.3cm}
\noindent
{\it Proof:}
The relative entropy distance reads
\begin{eqnarray*}
&&D\left(P_0(X)||P_\leftarrow (X|y)\right)=
\int p_0(x) (\ln p_0(x) -\ln p_\leftarrow (x|y)) dx\\ &=&
\int p_0(x) \Big(-\beta_1 H_1(x) dx-\ln \int e^{-\beta_1 H_1(x')} dx'\\&&+
\beta_1 (H_1(x)+H_{12}(x,y)) +\ln \int e^{-\beta_1 (H_1(x')+H_{12}(x',y))}  dx'\Big) dx\\ &=&
\int p_0(x) \beta_{1} H_{12}(x,y) dx+ \ln \frac{\int p_0(x) e^{-\beta_1(H_1(x)+H_{12}(x,y))} dx}{\int p_0(x) e^{-\beta_1 H_1(x)} dx}
\\&\leq& 
\int p_0(x) \beta H_{12}(x,y) dx\,.
\end{eqnarray*}
$\square$

\vspace{0.3cm}
\noindent
It is natural to ask whether one could also construct a limit where the fast system influences the  slow one without 
significant back action by assuming  $T_2\gg T_1$. However, the rescaling $T_2^{(n)}:=nT_2$ and $H_2^{(n)}(y):=n H_2(y)$ 
leads for $n\to \infty$ to  a conditional $P(y|x)=\exp(-H_2(y))/z$, i.e., $X$ and $Y$ become independent.

Note that there is a nice way to quantify action and back action in the above ``generalized equilibrium'' by a hypothetical sender/receiver
protocol. Assume a sender having access to system 1 adjusts his system to one value $x$ according to the marginal $p(x)$
in eq.~(\ref{MargEff}) above. Then the receiver observes values $y$ with probability $p_\rightarrow (y|x)$. 
His information about $X$ is 
given by the relative entropy distance between $P_\rightarrow(X,Y)=P(X)P_\rightarrow (Y|X)$ and $P(X)P_\rightarrow (Y)$.
Hence 
\begin{eqnarray*}
I_\rightarrow (X:Y)&=&\int p(x)  D\left(P_\rightarrow (Y|x)\Big\| \int P_\rightarrow (Y|x') p(x')dx'\right)  dx  \\
&\leq &\int p(x)p(x') D\left(P_\rightarrow (Y|x)||P(Y|x')\right) dxdx'\,,
\end{eqnarray*}
where we have used that relative entropy is convex \cite{cover}. 
If we define $I_\leftarrow (X:Y)$ in an analogue way, it follows that  the ``back action"-information
$I_\leftarrow (X:Y)$ tends to zero (in the limit $n\to \infty$). This is because calculations similar to the proof of Lemma~\ref{BackBound} show that 
$D(P(X|y)||P(X|y'))$ converge to zero for all $y,y'$. 
On the other hand, the ``forward information''
$I_\rightarrow (X:Y)$ converges to a non-zero value because the joint distribution on $X,Y$ obtained by the forward sender/receiver scenario
coincides exactly with the natural equilibrium $P$ defined by eqs.~(\ref{Pygx}) and (\ref{MargEff}) where we indeed have statistical dependences.

\section{Different aspects of simplicity}

\label{Sec:Aspects}

\subsection{Random walk on integers}

\label{Subsec:RandomWalk}

The second order Markov kernels are simple with respect to the following two criteria:
(1) The conditionals $p(x_j|x_1,\dots,x_{j-1})$ depend smoothly on $x_j$ and (2)  they depend
smoothly on $x_1,\dots,x_{j-1}$.
Now we will describe another aspect of simplicity that does not fit into these two categories.

Consider a random walk on $\Z$  (the set of integers) starting at position $0$.  In every step  we move either  one site to the left 
or one site to the right with probability $1/2$ each and stop after $n$ steps. Accordingly, we define the random variables
$X_1,\dots,X_n$ with values in $\Z$ describing the position after step $1,\dots,n$. The causal structure of the walk is certainly given
by the linear directed graph 
\begin{equation}\label{LineGraph}
X_1\rightarrow X_2\rightarrow \cdots \rightarrow X_n\,.
\end{equation}
 The corresponding conditionals for every variable, given its
parent node read:
\[
p(x_1=\pm 1)=1/2  \quad \quad p(x_j|x_{j-1})=\left\{ \begin{array}{ccc}  1/2 & \hbox{ for } & |x_j-x_{j-1}|=1 \\ 0 &  & \hbox{otherwise} \end{array}\right.  \,.
\]
The conditional independences entailed by the causal structure (\ref{LineGraph}) are also consistent
with the reverse causal hypothesis
\begin{equation}\label{ReverseLineGraph}
X_n \rightarrow X_{n-1} \rightarrow \cdots \rightarrow X_1\,.
\end{equation}
To see this, recall that we only need to check the Markov condition (Definition~\ref{MKC}). 
Given its parent $X_{j+1}$, every $X_j$ must be conditionally independent of all its non-descendants (except from its parent), i.e.,  the variables
$X_{j+2}$, $X_{j+3}$, $\dots$, $X_n$. Using the d-separation criterion in \cite{Pearl:00} one can easily show that this follows
from the Markov condition corresponding to the true causal structure (\ref{LineGraph}).
Due to eq.~(\ref{MarkovFac}) the joint probability then admits the factorization 
\[
p(x_1,\dots,x_n)=p(x_n)p(x_{n-1}|x_n)p(x_{n-2}|x_{n-1})\cdots p(x_1|x_2)\,.
\]
The conditionals $p(x_{j-1}|x_j)$ are, of course, also ``simple'' in the sense that they vanish for every pair $(x_{j-1},x_j)$ for which
$|x_{j-1} -x_j|\neq 1$. 
However, the conditionals are less simple in the sense that $p(x_{j-1}|x_j)$ {\it depends on} $j$.
To see this, assume that we are on position $\ell$ after $\ell$ steps. Then the position after step $\ell-1$ was definitely $\ell-1$.
In other words, the two cases $x_{j-1}-x_j=1$ and $x_{j-1}-x_j=-1$ are not equally likely for the backward time conditional\footnote{From the psychological point of view, it is remarkable that one is tempted to think that the backward-time conditional
would be the same for this example as the forward-time conditional.  This 
is consistent with a remark in the introduction: Our intuition seems to evaluate the simplicity of a model 
according to the simplicity of {\it causal} conditionals because we do not even recognize when a model is complex
in the converse direction.} 
and the bias depends on $j$.

The random walk represents 
another aspect of simplicity that is not taken into account in any of our inference rules proposed so far. 
It is  the
{\it simplicity of the dependence on the nodes} in the sense that
the function $j \mapsto p(x_j|x_{j-1})$ is simple since  it is even constant in $j$.
The ``physical'' reason is that the mechanism that determines the transition probabilities
is constant.

Due to the thermodynamic spirit of this paper it
is worth mentioning that 
the discussed time asymmetry is ``fading away'' after many  steps. 
This is because
\begin{equation}\label{RandomWalkEq}
p(x_j=\ell,x_{j-1}=m)=\left \{ \begin{array}{ccc} p(x_{j-1}=m)/2 &\hbox{ for } &  |\ell-m|=1 \\
                                           0 &  &\hbox{otherwise} \end{array}\right. \,.
\end{equation}
Since we have $p(x_{j-1})=m) \approx p(x_{j-1}=m\pm 1)$ for large $j$ 
the expression on the left hand side of Eq.~(\ref{RandomWalkEq}) becomes asymptotically symmetric
with respect to exchanging $\ell$ and $m$.
To obtain a strictly time-symmetric analogue, 
consider a random walk on a cycle consisting of  $N$ sites.
If the initial position is completely unknown, i.e., $p(x_1)=1/N$  for all $x_1\in \{0,1\dots,N-1\}$, 
the process is perfectly symmetric with respect to time inversion.

\subsection{Computational complexity in logical circuits}

\label{CompComp}

Here we want to describe an asymmetry with respect to
another notion of complexity, namely the complexity classes of
computer science.

First we consider a boolean function $f$ with
$n+m$ bits input and $k$ bits output.
Let $X=(X_1,\dots,X_n)$ be the vector of binary variables
that describe the first $n$ input bits and $Z=(Z_1,\dots,Z_m)$ be
the vector for the last $m$ input bits. Let furthermore $Y:=(Y_1,\dots,Y_k)$
describe the output. Now we interpret
$X$ as the {\it cause}, $Y$ as the {\it effect}
and $Z$ as a {\it noise}
variable that makes the causal mechanism probabilistic.
We will assume that the total input (including ``cause'' and ``noise'') is obtained by  statistically
independent initialization of the bits.

The following statement is almost obvious:

\begin{Observation}[approximating P(effect$|$cause)
is efficient]${}$\\
\label{PEffC}
 Given a string  $d$ including

\begin{enumerate}
\item a description of
 a boolean circuit
in terms of elementary gates like AND, NAND, OR, NOR, NOT
that computes $f$ and

\item a description of a product probability distribution
for  $Z$.
\end{enumerate}

Given some $\epsilon$ with $\epsilon = 1/{\rm poly} (|d|)$ and some constant
$c\in (0,1)$. The problem to decide for a given pair $x,y$
whether
\[
p(y|x) \geq c+\epsilon \hspace{1cm}\hbox{ or } \hspace{1cm}
 p(y|x) \leq c-\epsilon
\]
is in BPP (``Bounded-error, Probabilistic, Polynomial time''  \cite{Papa}). In other words, there is a probabilitistic algorithm whose running time
increases only polynomial in $|d|$ solving the above decision problem
such that the error probability is smaller than some
previously specified constant $\delta >0$.
\end{Observation}

\noindent
The ``algorithm'' for this decision problem is already given by
setting the input to $x$,
randomizing the noise variable $Z$ according to the given
distribution, simulating the boolean circuit and counting the number
of runs with output $y$.

Is should be noted, however, that an exact computation of $P(y|x)$
is not possible in any efficient way provided that the complexity classes $\# P$ (``sharp P'')
and BPP
do not coincide.
To see this, we set $n=0$ and $k=1$,  i.e.,
the binary variable
$Y$ is only a function of the noise
$Z$. Let the values of the noise variable be uniformly distributed, i.e.,
$p(z)=1/2^m$ for all $z \in \{0,1\}^m$.
Then $p(y|x)=p(y)$ is, up to the constant $1/2^m$ simply the
number of inputs $z$ for which $f(z)=1$. The problem to count the number of
satisfying inputs for a boolean function (given in so-called conjunctive normal form)
\[
f:\{0,1\}^m \rightarrow\{0,1\}
\]
is complete for the complexity class $\#$P.
This class
is believed to contain extremely
hard computational problems \cite{Papa}.
However, the hardness of giving {\it exact} solutions
is probably of minor relevance and we will
 now consider {\it approximative} solutions.

We will see that  $P(\rm{cause}|\rm{effect})$ is even hard to compute {\it approximately}:

\begin{Theorem}[approximating P(cause$|$effect) is NP-hard]${}$\\
Let the assumptions and definitions be as in Observation~\ref{PEffC} with general $n,m,k\in  \N_0$
and $p(x)$ be some product distribution.
Then the problem to decide whether 
$p(x|y) \geq 2/3$ or $p(x|y) \leq 1/2$ for a given pair $(x,y)\in \{0,1\}^n\times \{0,1\}^k$ is NP-hard. 
\end{Theorem}

\vspace{0.3cm}
\noindent
{\it Proof:}
NP-hardness can even be proved for the special instance $m=0$, i.e., without introducing
a noise variable $Z$.   This shows that the general problem contains NP.
Let $g:\{0,1\}^n \rightarrow \{0,1\}$ be a boolean function.
To decide whether there is a binary string $x$ with $g(x)=1$ is known to
be NP-complete \cite{Papa}.
We chose $g$ such that $g(0,\dots,0)=0$.
It is clear that the  restriction to this class of functions $g$
remains NP-complete. Then we define a function $f$ by $f(x)=g(x) \vee  h(x)$
with $h(x)=1$ for $x=(0,\dots,0)$ and $h(x)=0$ otherwise.
Let now the distribution of $x$ be uniform and
consider $p(x=(0,\dots,0)|y=1)$.
If $g$ gas no satisfying
input $x$ we have $p(x=(0,\dots,0)|y=1)=1$ since $x=(0,\dots,0)$ is
the only satisfying input for $f$. If $g$ has a satisfying input,
$f$ has at least two satisfying inputs and
 $p(x=(0,\dots,0)|y=1)\leq 1/2$.
$\Box$

\vspace{0.3cm}
To better understand the reason for the asymmetry between the complexity
of computing $P(\rm{effect}|\rm{cause})$  and $P(\rm{cause}|\rm{effect})$
we extend the boolean function $f$ to a {\it bijective} function $F$.

The Toffoli gate \cite{Toffoli} provides a useful method to simulate
conventional boolean circuits by reversible ones.
TOFFOLI is a gate with three inputs $a,b,c$  and three outputs
$a',b',c'$ such that $a'=a$, $b'=b$ and $c'=c\oplus (a\cap b)$ where
$\oplus$ denotes the exclusive or (``XOR'').
In words, the third bit $c$ is inverted if and only if $a$ and $b$ are true.
TOFFOLI can simulate NAND by setting the third input to $c=1$.
Then we have  $c'=\overline{a\cap b}$ and the outputs $a',b'$ can be
ignored (note that the existence of ``useless'' output (``data garbage'') and the need
for adjusting certain input bits to fixed values is characteristic
for reversible computation). 

Since NAND is universal and
gates
like AND, OR, NOR, NOT can be simulated using a small number of NAND
gates we
can simulate every given boolean circuit  with TOFFOLI gates efficiently. Furthermore,
a corresponding reversible circuit can be found efficiently by
substituting every single gate with some TOFFOLI gates.

\begin{figure}\label{Schaltkreise}
\centerline{
\epsfbox[0 0 132 177]{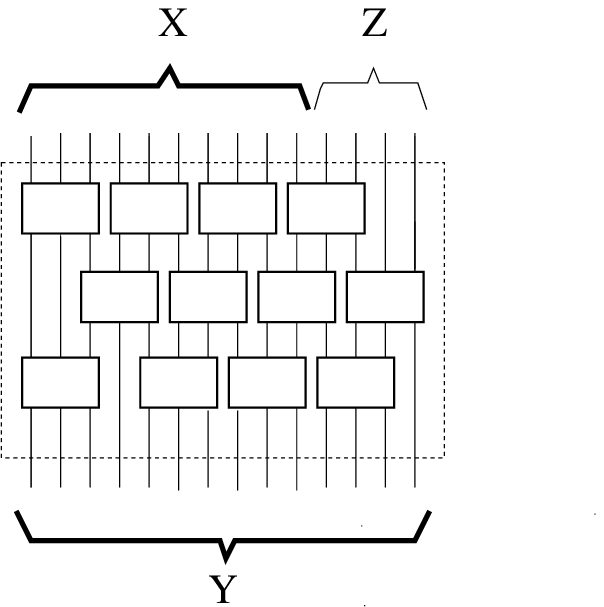} \hspace{2cm}
\epsfbox[0 0 132 177]{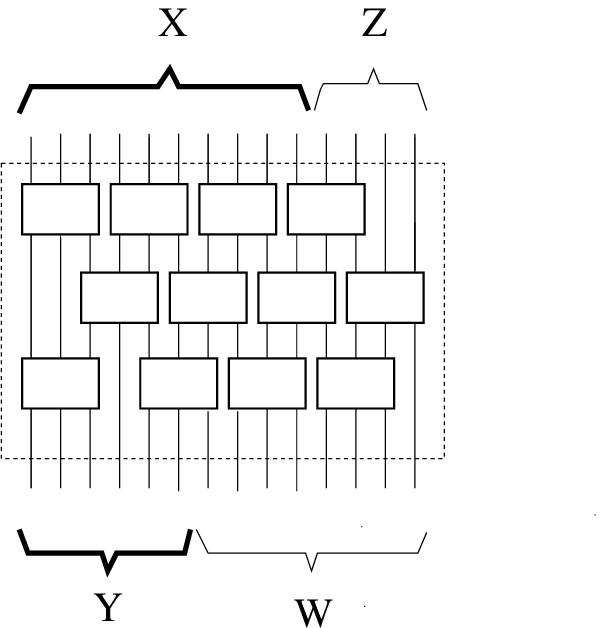}
}
\caption{{\small
Reversible network as a model for a cause-effect relation.
Left: The ``cause variable $X$'' and the noise variable $Z$
determine jointly the effect variable $Y$ and vice versa.
Both conditionals $p(y|x)$ and $p(x|y)$
are efficiently computable.
Right: The effect variable $Y$ does not completely determine the cause variable
$X$ and the noise $Z$. The conditional
$p(x|y)$ is in general not efficiently computable, but $p(y|x)$ is.
}}
\end{figure}

This yields an algorithm to extend $f$ (having $n+m$ input bits and
$k$ outputs)
to a bijective boolean function
$F$ with $\tilde{n}=n+m+r$ inputs and $k+l=\tilde{n}$ outputs
(described by $Y,W$)
such that
for some additional $r$-bit string $v$ the restriction of
$F(x,z,v)$ to the first $k$ output bits coincides with $f(x,z)$.
We may without loss of generality consider the ancilla variables $V$
as additional noise
variables since we can specify the corresponding distributions
such that the ``noise'' variable always attains the same value. 
Hence we  obtain a boolean function $F$ with $n+m$ input bits and
$k+l=m+n$ output bits such that the conditional
probabilities for the output $y$ given the input $x$
coincide with the probabilities $p(y|x)$ generated by
the function $f$. We have then simulated the causal effect from $X$ to $Y$
by a completely reversible process using a noise variable $Z$ and
restricting the output $Y,W$  to $Y$ (see Fig.~\ref{Schaltkreise}).

It is important to note that the inverse function $F^{-1}$
can be computed efficiently: every TOFFOLI gate is its own inverse.
We can therefore simulate the circuit in backward direction.
In such a setup both $p(y|x)$ and
$p(x|y)$ are efficiently computable
provided that
$Y$ is the {\it complete} output.
This is because we can compute
the complete input $(x',z) =F^{-1}(y)$ from $y$.
Then we know that  $p(x|y)=1$ for  $x=x'$ and $p(x|y)=0$ otherwise. 

It should be emphasized that {\it local} reversibility
of the network is essential, i.e.,
it is not sufficient that the computed function $F$ is {\it bijective}.
In order to compute $F^{-1}$ efficiently we have inverted each single
gate.\footnote{An example for a bijective function whose inverse
is believed to be not
efficiently computable (because it is not {\it locally} invertible)
is $f(n)=(a^n)$ $mod\, b$ where $a$ and $b$ are chosen appropriately and $n \in \{0,\dots,b-1\}$. The security of the
crypto-system RSA relies on
the assumption that the inverse of this function is hard to find.}

We conclude: If $y$ is the complete output of
a locally reversible circuit we can compute $p(x|y)$ efficiently.
In other words, using the complete effect of the cause, we would
be able to compute
$P(\hbox{cause}|\hbox{effect})$
efficiently, no matter whether we have probabilistic
causality where $y$ is additionally influenced by a latent variable.

It has been argued \cite{Landauer:61,BennettThermoReview} that logically irreversible functions
lead to energy dissipation and thus
thermodynamically reversible computation
is only possible by computing only  reversible functions
\cite{Bennett:73,Toffoli}. For this reason, the ``garbage''  bits $w$ directly correspond to heat generation. 
The following theorem provides an upper bound on the complexity of computing the backwards conditional in terms of the number of  garbage bits.

\begin{Theorem}[complexity of P(cause$|$effect) and thermodynamics]${}$\\
The decision whether $p(x|y)>2/3$ or $p(x|y)<1/3$ 
requires at most the following steps: 
(1) the estimation of $p(x,y)$, (2)
$2^\ell$ queries of $F^{-1}$  
 when $\ell$ is the number of garbage bits, and (3) estimating  the probability of $2^\ell$ input  strings.  
\end{Theorem}

\vspace{0.3cm}
\noindent
{\it Proof:} using
\[
p(x|y)=\frac{p(y|x)p(x)}{p(y)}=\frac{p(y|x)p(x)}{\sum_w p(F^{-1}(y,w))}\,,
\]
the statement is obvious  since  every term  $p(F^{-1}(y,w))$ can be computed using one query of $F^{-1}$ 
followed  by the estimation of $p(F^{-1}(y,w))$.
$\Box$ 

\vspace{0.3cm}
Using the reversible embedding, it becomes obvious that
the asymmetry between cause and effect has
been put in {\it by assumption}:
We have postulated that all input bits are statistically independent.
We could think of the time-inverted scenario
where
$x,z$ is distributed according to some probability distribution $P$ having
the property that the distribution of $y,w$ is a product measure.
Then we could efficiently compute $p(x|y)$ applying the
method in Observation~\ref{PEffC} and obtain
an efficient  simulation of the time-reversed circuit.
Obviously, such a scenario is unlikely unless
{\it we have calculated} how to randomize the input such that a
product distribution of the output is obtained.  

This asymmetry becomes a more physical interpretation if 
we think of the bits as states of physical systems 
that have never been interacting before some time $t_0$. After they interact, a collective dynamics (represented by the circuit)
creates stochastic dependences between initially independent systems.

\section{Common root of the asymmetries}

\label{Sec:CommonRoot}

This section provides a unified view on the origin of the following
facts:

\noindent
(1) the asymmetry of the computational complexity for the circuit in the preceding subsection

\noindent
(2) the asymmetry of the causal Markov condition under inverting arrows 

\noindent
(3) the asymmetry of models with second order Markov kernels

The common root is the
tendency of our environment to subject a system to interactions with an abundance of
other physical systems that are initially uncorrelated with the former rather than being finally uncorrelated
(this has already been described for the logical circuit).
It is clear, that this tendency is linked to other asymmetries between past and future:
We see a scene happening in front of our eyes 
shortly {\it after} it has happened because the photons absorbed by the eyes 
have obtained correlations with the objects at which they were reflected. 
The physical state of the light beam was uncorrelated with the object {\it before} it interacted with the latter but correlated {\it afterwards}.
This is consistent with Reichenbach's principle, saying that statistical dependences have to be explained by interactions in the past
but not by interactions going to happen in the future. Hence, some evident asymmetries between past and future are related to the principle of the common cause \cite{penrose62direction}.

To discuss these links we will use 
{\it classical microphysical} toy models:

\noindent
(1)  Different random variables represent the state of
a different physical system or the state of the same system at a
different time. This means that the value set of the variable is identified with the space of {\it pure} states and the set of 
measures on the value set is the
spaces of {\it mixed} states. 

\noindent
(2) The space of pure states of a composed  system is given by the Cartesian product
of the spaces of the constituents.
 
\noindent
(3) A physical process of a closed physical system is a bijective map on its set of pure states.

\noindent
(4) A physical process of an open physical system is a bijective map on the Cartesian product of the set of pure states
of the system under consideration and the set of pure states of an additional system, called the environment. 

Our classical microphysical models are discrete, i.e., one may interpret them as quantum systems whose density operators
are restricted to those being diagonal with respect to some fixed basis.

\subsection{Microphysical model for common causes and common effects}

The statistical asymmetry between a causal fork and a causal collider (see Fig.~\ref{FC}) 
 is only  the simplest case for   the 
asymmetry of  the causal  Markov condition with respect  to reversing arrows, but  
the crucial idea can already be seen from this case.

If $Z$ is the common cause of $X$ and $Y$ we recall
\begin{equation}\label{fork}
X \independent Y \,| Z \,\,\,\hbox{ but }\,\,\, X  \not\independent Y\,,
\end{equation}
where $\independent$ denotes independence and $.\independent .| .$ is conditional independence.
If $Z$ is the common effect of two (causally) independent causes we
have in the generic case
\begin{equation}\label{collider}
X \not\independent Y |\,Z \,\,\,\hbox{  but }\,\,\, X \independent Y
\end{equation}
Already Reichenbach \cite{Reichenbach}  
discussed this asymmetry in the context
of mixing processes in interacting  dynamical systems.
The following subsection is not far from Reichenbach's idea. However, to 
describe the common thermodynamic root of the asymmetry between (\ref{fork}) and (\ref{collider}) on the one hand and 
the asymmetry postulated by the principle of plausible Markov  kernels  on the other hand (Subsection~\ref{Subsec:Shape}) we 
have chosen a class of models 
that is appropriate to discuss both types of asymmetries.

\vspace{0.3cm}
\noindent
{\it Causal fork (common cause):}
Let $P(X,Y,Z)$ be a jointy distribution generated by a causal structure where $Z$ is the common cause of
$X$ and $Y$. 
We construct a bijective 
process acting simultaneously on $6$ systems
\[
S_Z\times S_X \times S_{NX} \times S'_Z \times S_Y\times S_{NY}\,.
\]
Their role is as follows.
The initial state of $S_Z$ represents the variable $Z$ and the final states 
 of $S_X$ and $S_Y$
(after some bijective process has acted jointly on
the $6$ systems) represent the variables $X$ and $Y$, respectively.
The time order guarantees that $Z$ can only be a cause and not an effect of $X$ and $Y$. 
 Systems $S_{NX}$ and $S_{NY}$
represent background noise that prevents $X$ and $Y$ from being deterministic functions of
$Z$ (see remarks after Definition~\ref{MKC}). The role of $S'_Z$ is a bit more subtle and is easier to explain after 
the process has been described. Let $P$ be a product distribution on
$S_Z\times S_X\times S_Y \times S_{NX}\times S_{NY}$ and let $S'_Z$ be in an arbitrary pure state.
Then we construct a process consisting of two steps. 

\vspace{0.3cm}
\noindent
(Step 1) Apply a bijective map $F_Z$ on
\[
S_Z\times S'_Z\,.
\]

\vspace{0.3cm}
\noindent
(Step 2)
Apply bijective maps  $F_X$ and $F_Y$ on
\[
S_Z\times S_X\times S_{NX} \quad \hbox{ and } \quad S'_Z\times S_Y\times S_{NY}\,,
\] 
respectively.
Note that 
$F_Z$ distributes the information contained in $Z$ such that it (or at least part of it) is afterwards available on  $S_Z$ and $S'_Z$. 
This ``broadcasting'' of information into two components ensures that $S_Z$ can have an effect on both $S_X$ and $S_Y$ 
even though a direct interaction between $S_X$ and $S_Y$ is avoided. 
If $F_X$ and $F_Y$ both would act (one after another) on $S_Z$ 
we could not exclude information transfer between them in contradiction to our causal model being a fork.

It is easy to show that processes of the above kind generate a distribution with $X\independent Y\,|Z$. To show that every
joint distribution on $X,Y,Z$ satisfying $p(x,y,z)=p(z)p(x|z)P(y|z)$ can be generated by a process of this type, 
we  assume that $S_Z'$ starts in a state with zero entropy and $F_Z$ copies the value of $z$ so that it is afterwards available on both systems $S_Z$ 
and $S_Z'$. Using appropriate ``noise systems'' $S_{NX}$ and $S_{NY}$ the maps $F_X$ and $F_Y$ can certainly generate any desired transition matrices $p(x|z)$ and
$p(y|z)$, respectively. 

\vspace{0.5cm}

\noindent {\it Collider (common effect):} Here we consider the same
6 systems and the same bijections, but in time-reversed order.
Let $P$ be a product distribution on
$S_X\times S_Y\times S_Z \times S_{NX}\times S_{NY} \times S'_Z$.
 Then implement
$F_X$ and $F_Y$ as above and $F_Z$ afterwards. 
Note that $X$ influences the final state of $S_Z$ via first influencing its intermediate state
(between steps 1 and 2)
and $F_Y$ influences the  final state of $S_Z$ via first influencing $S_Z'$.
Then we have $X \independent Y$ by assumption, but
not necessarily $X \independent Y | Z$.

The backward time  version of scenario 1 would be
the following. The joint system is {\it initially} correlated in a way
that ensures that the application of  $F_X,F_Y$ and $F_Z$ makes them
statistically independent. This would be a rather contrived
situation. We do not claim 
that every physical system is ``initially'' uncorrelated from its
environment. 
The essential point making the backward scenario unlikely is
that the correlations are exactly such that the dynamics
resolves them into a product state. Since the dynamics is bijective this is unlikely to  be the case
unless initial state  and the process are adjusted  \cite{TasakiGaspard,GaspardChaos} 
to  each other \footnote{This would mean that  initial state and dynamics have  
{\it algorithmic information} in common. According to the 
algorithmic Markov condition postulated in \cite{Algorithmic} this requires
a causal connection between these two ``objects''.}. 

To be consistent with our model class, we rephrase this asymmetry as follows.

\begin{Postulate}[Arrow of time in a closed system]${}$\\
Let $\cX_j$ denote the state space of system $j$.  
Let the dynamics after some time $t$ 
be given by a bijective map
\[
F: \cX\rightarrow \cX
\]
with
\[
\cX:=\times_{j=1}^k \cX_j\,.
\]
Let $P$ be a probability distribution on $\cX$ that formalizes the initial 
statistical state of the system and $P\circ F$ denote its final state. 
 
If $k$ is large, it is unlikely that $P\circ F$ is a product state but $P$ is not (unless $F$ has been  designed ``by hand'' in order to
transform the non-product state into a product state). 
The reverse scenario, that $P$ is a product state and $P\circ F$ is not, happens quite often. 
\end{Postulate}

A typical permutation of $k$ tuples in the $k$-fold Cartesian product creates dependences if the initial distribution 
is a product measure whose entropy is not maximal. This can be considered as a model for increasing correlations 
being the typical situation in
{\it closed} systems. 
This is certainly directly connected with the usual arrow of time in statistical physics
where interactions between particles lead typically to an increase of coarse-grained entropy
(cp. e.g. \cite{Gibbs,penrose62direction,Lebowitz}).

In open systems, however, we have to take into account the following effect:
The restriction of a probability distribution on a Cartesian product to  a small fraction of subsystem is typically close to a maximal entropy distribution
(hence a product measure) even though the distribution itself may be far away from a product distribution.
In quantum systems, we have even the stronger statement that the restriction of a typical  {\it pure} many-particle 
state is so strongly entangled that its restriction to a small fraction of subsystems is almost the maximum entropy state \cite{WinterMix}.
Therefore, it was important for the justification of our way of reasoning that we considered maps on {\it closed} systems by
taking the environment explicitly into account (in form of $S_{NX}$ and $S_{NY}$). Otherwise we could not justify the remark that
increase of dependences is more typical than resolving dependences.

\subsection{Asymmetry in the shape of conditionals}

\label{Subsec:Shape}

\begin{figure}
\centerline{\includegraphics[height=6cm]{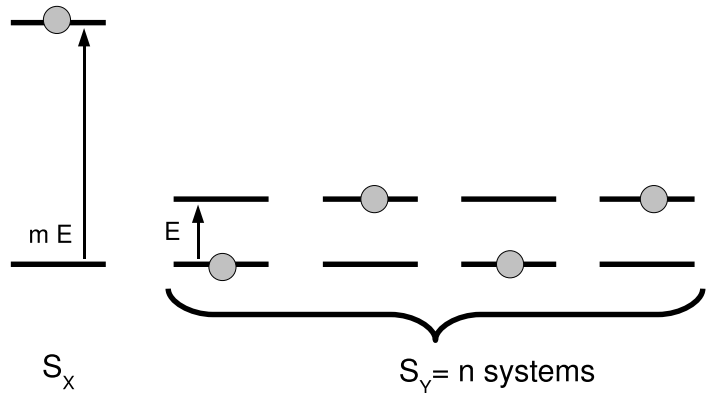}}
\caption{{\small
Relaxation process with $n+1$ weakly interacting two-level systems. 
A random process distributes the initial energy over the joint system.
The initial energy of $S_X$ probabilistically influences the
the final energy of $S_Y$. This provides a model for a binary variable
influencing an almost continuous one.
On the other hand,
the initial energy of  $S_Y$ influences the final energy of $S_X$. Then an almost  continuous variable 
effects a binary variable.\label{SpinRelax}}}
\end{figure}

Now we describe a scenario where a mixing process of a simple physical system reproduces
our second order Markov kernels under appropriate conditions.
Let system $S_X$ 
be a classical two-level system with energy gap $E_X=1$ and system $S_Y$ consist of
a large number $n$ two-level systems with energy gap $E_Y=mE_X=m$ as shown in
Fig.~\ref{SpinRelax}.

We assume that $m$ grows asymptotically proportional
to $\sqrt{n}$, i.e., $m_n=c_n \sqrt{n}$ with $c_n\to c$. 
Moreover, the initial joint  distribution of
the  $n+1$ two level systems is  a product distribution where
the upper level of $S_X$ is occupied with  
probability $r$ and the upper level of each system in $S_Y$ with probability $q$.
Then we assume that a weak interaction drives a mixing process 
on the joint state space $\{0,1\}^{n+1}$
that randomly permutes levels with the same total energy.

We define a binary variable $X$ describing the state of $S_X$
and a variable  $Y$ that is asymptotically continuous for $n\to\infty$. 
It's values are given by
\[ 
y:=\frac{\ell-nq}{\sqrt{n}}\,,
\]
where $\ell=0,\dots,n$ denotes  the total energy of $S_Y$.

Let $X_i,Y_i$ and $X_f,Y_f$ refer to the initial and final states, respectively.
Certainly, $X_i$ and  $Y_i$  influence $X_f$ and $Y_f$. However, we will focus
on two variables only and, for instance, say that $X_i$ influences $Y_f$. Then  $Y_i$ is considered
as a noise adding further indeterminism to the causal influence. Hence, the
 mixing process can be considered as a model for the causal structures
$X_i\rightarrow Y_f$ and $Y_i\rightarrow X_f$ at the same time. 

We will discuss
the process for different choices of $q$ and  $r$ 
in  the limit $n\to\infty$
and show that only the following  three cases occur:

\noindent
(1) the joint distribution between initial and final variable
does not have a  second order model in any direction,  neither the temporal nor the time reversed one.

\noindent
(2)  it has a second order model in both directions. 

\noindent
(3) a second order  model exists  only in the temporal  direction.

It will become obvious that the only time asymmetry in the below scenario is  that we assume 
statistically independent two-level systems as initial condition instead of imposing  independence as a final condition.

We introduce the following notation. The joint density $p(y_f,x_i)$ is said to be  in $S_{X_i\rightarrow Y_f}$ if
$p(x_f|x_i)$ and $p(x_i)$ are asymptotically second order Markov kernels in the sense of Definition~\ref{SC}
or can be approximated by these  type of conditionals. 

By  the usual central  limit theorem,
$p(y_i)$ is asymptotically Gaussian with mean zero and variance $q(1-q)$.
In the below discussion, we will always refer to  the asymptotical case 
unless the converse is explicitly stated.
To compute the final distributions
we have to distinguish between different regimes of $q$ (which corresponds  to different initial 
temperatures of $S_Y$).

\newpage
%\vspace{0.3cm}
\noindent
{\bf Finite temperature:}\\
Let $q\neq 1/2$. We discuss only $q<1/2$ since  $q>1/2$ (``temperature inversion'')  
is similar with exchanging the role  of upper and lower levels.

\vspace{0.3cm}
\noindent
{\it X  influences Y:}
Asymptotically, $x_f$ will always be zero. This is intuitively clear because 
the energy gap of $S_X$ tends to infinity.
More formal  arguments can be constructed in analogy to the derivations below.
Therefore the whole total energy is finally in $S_Y$ and $p(y_f|x_i=0)$ is Gaussian with mean zero and
variance $q(1-q)$.

If $S_X$ starts in its  upper state instead, the total energy  is shifted by $c_n\sqrt{n}$
and $p(y_f|x_i=1)$ therefore is Gaussian with mean $c$ and variance $q(1-q)$.
This  shows that $p(y_f|x_i)$ is second order. Since $p(x_i)$ is  trivially  second order we obtain a joint distribution
$p(y_f,x_i)$ in the class $S_{X_i\rightarrow Y_f}$.
Since $p(y_f)$  is a Gaussian mixture, it cannot be in $S_{Y_f\rightarrow X_i}$.

\vspace{0.3cm}
\noindent
{\it Y influences X:}
In fact there is no influence because $p(x_f=0)=1$. 
The joint distribution $p(y_i,x_f)$
 is in $S_{Y_i\rightarrow X_f}$ and $S_{X_f \rightarrow X_i}$ 
because $p(y_i,x_f)=p(y_i)p(x_f)$ and $p(y_i)$ is Gaussian and $p(x_f)$ is second order anyway. 

\vspace{0.3cm}
\noindent
{\bf Infinite temperature:}
Let $q=1/2$. Then $S_X$ does not necessarily end up in its lower level.
We first compute $p(x_f|x_i=0,y_i)$ for the case of finite $n$.
If $S_X$ ends  up in its lower state the initial total energy $\ell_i$ is distributed among the subsystems of 
$S_Y$, otherwise 
we only have to distribute the energy  $\ell_i-m$.
The ratio between the number  of combinations for both cases  
provides the ratio between the probability to find $S_X$ in its upper or lower level  after  the mixing:
\begin{equation}\label{combin}
\frac{p(x_f=1|x_i=0,y_i)}{p(x_f=0|x_i,y_i)}=\frac{\ell_i(\ell_i-1)\cdots (\ell_i-m_n+1)}{ (n-\ell_i+1)(n-\ell_i+2) \cdots (n-\ell_i+m_n)}\,. 
\end{equation}
Taking the logarithm of the right hand side and using  $\ell_i=\sqrt{n}y_i+n/2$ yields
after same algebra 
\[
\sum_{j=0}^{m_n-1} \ln \left(1+ \frac{2\sqrt{n} y_i-2j-1}{n/2-\sqrt{n}y_i+j+1}\right)=:\sum_{j=0}^{m_n-1} \ln (1+W_j)\,.
\]
Every $W_j$ tends to zero with $O(1/\sqrt{n})$ and  the  sum consists of $O(\sqrt{n})$ terms. Due to
$\ln (1+W_j)=W_j+O(W_j^2)$ we thus have
\begin{eqnarray*}
\lim_{n \to \infty} \sum_{j=0}^{m_n-1} \ln (1+W_j)&=&\lim_{n\to\infty} 
\sum_{j=0}^{m_n-1} W_j=\lim_{n\to\infty} \sum_{j=0}^{m_n-1}  \left(\frac{4y_i}{\sqrt{n}} -\frac{4j}{n}\right)\\
&=& 4cy_i -\int_0^c4x dx=2c(2y_i -c)\,.
\end{eqnarray*} 
The second equation reduces the expression to the 
asymptotically relevant terms
 and the third step holds 
because $m_n$ grows with $c\sqrt{n}$.
Hence
\[
\frac{p(x_f=1|x_i=0,y_i)}{p(x_f=0|x_i,y_i)}  = e^{2c(2y_i-c)}\,,
\]
i.e.,
\begin{equation}\label{distr1}
p(x_f|x_i=0,y_i)= \frac{1}{2}\Big( 1\pm \tanh[2c(2y_i-c)]\Big) \,,
\end{equation}
where the signs $+,-$ correspond to $x_f=1,0$,   respectively.
For $x_i=1$ the initial total energy is $\ell_i+m$ instead  of $\ell_i$  and $y_i$ is thus replaced with $y_i+c$: 
\begin{equation}\label{distr2}
 p(x_f|x_i=1,y_i)= \frac{1}{2}\Big( 1\pm \tanh[2c(2y_i+c)]\Big)\,.
\end{equation}

\vspace{0.3cm}
\noindent
{\it Y influences X:}
For $r=0,1$ the conditional $p(x_f|y_i)$  is given by (\ref{distr1}) or  (\ref{distr2}), respectively.
Hence
$p(x_f,y_i)$  is in $S_{Y_i\rightarrow X_f}$ because $p(y_i)$ is a Gaussian with mean zero and variance $1/4$, i.e.,  
\[
p(y_i)= \sqrt{\frac{2}{\pi}} e^{-2y_i^2}\,. 
\]
To see that $p(x_f,y_i)$  is not in $S_{X_f\rightarrow Y_i}$ we recall that only for the trivial cases
the joint distribution is second order in both directions (see Section~\ref{PlMK}).

For $r\neq 0,1$ we obtain a  mixture of the conditionals (\ref{distr1}) and (\ref{distr2}), which is no  longer of second order
and hence $p(x_f,y_i)$ is not in $S_{Y_i\rightarrow X_f}$. To see that it  is not in $S_{X_f\rightarrow Y_i}$ either,
we observe
\begin{eqnarray*}
p\big(y_i,x_f=1\big)&=&p\big(x_f=1|y_i\big)\,p(y_i)\\
&=&\frac{1}{2}\Big(1+\tanh[2c(2y_i+c)]+\tanh[2c(2y_i-c)]\Big) \,\sqrt{\frac{2}{\pi}} e^{-2y_i^2}\,,
\end{eqnarray*}
which is not proportional to a Gaussian as it should be.

\vspace{0.3cm}
\noindent
{\it X influences Y:}
To compute $p(y_f|x_i)$ we observe 
\begin{eqnarray*}
p(y_f|x_i)=\sum_{y_i,x_f} p(y_f,x_f|x_i,y_i)p(y_i)\,,
\end{eqnarray*}
because $X_i$ and $Y_i$ are  independent.
Since
$y_i=y_f+c(x_f-x_i)$ we  obtain
\begin{eqnarray*}
p(y_f|x_i)&=&\sum_{x_f} p\Big(x_f\Big|x_i,\,y_i=y_f+c(x_f-x_i)\Big)p\Big(y_i=y_f+c(x_f-x_i)\Big)\\
&= &  \frac{1}{2}\Big(1+\tanh[2c(2y_f-2cx_i+c)]\Big) \,\sqrt{\frac{2}{\pi}} e^{-2(y_f+c(1-x_i))^2} \\
&+& \frac{1}{2}\Big(1-\tanh[2c(2y_f-2cx_i-c)]\Big) \,\sqrt{\frac{2}{\pi}} e^{-2(y_f-cx_i)^2}\,.
\end{eqnarray*}  
Plotting $p(y_f|x_i)$ for fixed $x_i$ shows that it is not Gaussian for fixed $x_i$  unless $c=0$. 
Likewise, $p(y_f)$ is not Gaussian. Hence $p(y_f,x_i)$ neither is in $S_{X_i\rightarrow Y_f}$ 
nor in  $S_{Y_f\rightarrow X_i}$. 

\vspace{0.3cm}
Spin systems are actually quantum systems. In order to further support the general idea we want
to sketch a corresponding quantum scenario.

Let $S_X$ and $S_Y$ be described by the Hilbert spaces $\cH_X:=\C^2$ and $\cH_Y:=(\C^2)^n$, respectively
and $S_X$  start in its lover level $|0\rangle$.
We assume that $S_Y$ starts in an eigenstate 
$|\psi\rangle\in \cH_Y$ of the total energy with eigenvalue $\ell$.

Now we discuss what a typical {\it energy-conserving unitary map} on $\cH_X\otimes \cH_Y$ does. 
The space of states with total energy $\ell$ splits up into the space 
\[
\cH_0:=|0\rangle \otimes \cG_{\ell}\,,
\]
where $G_{\ell}$ consists of all states in $\cH_Y$ having total spin $\ell$, and
\[
\cH_1:=|1\rangle \otimes \cG_{\ell -m}\,.
\]
Obviously, the  quotient of the dimensions coincides with the quotient of the number of combinations
given by the right hand side  of eq.~(\ref{combin}). 

After the unitary process has been applied, we have a state of the form
\[
|0\rangle \otimes |\psi_0\rangle +|1\rangle \otimes |\psi_1\rangle\,,
\]
with $|\psi_0\rangle\in \cG_{\ell}$  and $|\psi_1\rangle \in \cG_{\ell-m}$.  
The probability that $S_X$ is found in its upper level is then given by $\| |\psi_1\rangle \|^2$.
The following lemma shows that in high dimensions almost every state in $\cH_0\oplus \cH_1$ 
has the property that $\| |\psi_1\rangle\|^2/\| |\psi_0\rangle \|^2$ is close to the quotient of the dimensions of $\cH_1$ and $\cH_0$.

\begin{Lemma}\label{Lemma:Di}
Let $\cH_1$ and $\cH_2$ be two Hilbert spaces of dimension $d_1$ and
$d_2$, respectively. Let $|\psi\rangle :=|\psi_1\rangle \oplus
|\psi_2\rangle$ be a randomly chosen vector according to the Haar
measure of $SU(d_1+d_2)$. Then,  for every $\eta>0$, the probability that
$|\|\psi_1\rangle \|^2/\|\psi_2\rangle\|^2 -d_1/d_2|>\eta$ tends to
zero for $d_1+d_2 \to \infty$.
\end{Lemma}

\vspace{0.3cm}
\noindent
{\it Sketch of the proof:} Define a function $f$ on $\cH$ by
\[
f(\psi):= \langle \psi | P_1| \psi \rangle\,,
\]
where $P_1$ is the projector onto $\cH_1$. 
The function $f$ is  Lipschitz-constant with $L=2$. 
It is easy to show that
the average of $f$ over $SU(d_1+d_2)$ is $d_1/d_2$. Otherwise the
Haar measure would not be invariant with respect to permutations of
basis vectors. Then the Lemma follows from Levis Lemma
\cite{WinterMix}: 
Given a Lipschitz continuous function $f$ on a unit sphere,
the volume of the region where $f$ is not close to its average can be bounded from above in terms of  $L$.
For growing dimension  the  bound tends to zero. 
$\Box$ 

\vspace{0.3cm}
\noindent
This shows that the probability to find $S_X$ finally in its upper level depends on the same way on $\ell$
as in the  classical scenario. Hence we reproduce the second  order Markov kernel
in eq.~(\ref{distr1}).

However, the difference is that the outcome 
for $x_f$-measurements even is probabilistic when one specific {\it pure } initial state and 
one typical unitary is considered (without requiring any further randomness).

\section{Relation to non-equilibrium thermodynamics}

\label{NonEq}

Explaining  the postulated asymmetries via mixing  processes, as we have done repeatedly,
suggests to consider the topics discussed by this paper as part of {\it non-}equilibrium thermodynamics.
However, a priori, it is not clear     
whether there could also be statistical asymmetries between cause and effect in thermal {\it equilibrium}.

We first consider the difference between past and future in stochastic processes $(X_t)_{t\in \Z}$
where the answer is negative.
If  the factorization  
\[
P(X_t,X_{t-1})=P(X_{t-1})P(X_t|X_{t-1})
\]
leads to simpler terms  than 
the factorization into $P(X_t)P(X_{t-1}|X_t)$ in any sense then we must have a violation of the symmetry
\begin{equation}\label{DB}
P(X_t=r,X_{t-1}=s)= P(X_t=s,X_{t-1}=r)\,.
\end{equation}
If the (possibly vector-valued) variable $X_t$ describes the state of a physical system in phase space at  time $t$ 
eq.~(\ref{DB}) is just another formulation of the so-called {\it detailed balance} condition which is known to hold in Gibbs equilibrium
\cite{Tolman}, but not in non-equilibrium steady states  \cite{Klein}. 
This shows that statistical asymmetries between past and future require non-equilibrium states.
In the  literature, such asymmetries have been discussed  for various types of 
non-equilibrium steady states, e.g. \cite{Gaspard,GapardSt} as well as the relation to 
thermodynamic irreversibily.

To explore the importance of non-equilibrium for the models discussed in this paper, 
we first consider an extremely simplified quantum model of the dynamics in the Stern-Gerlach experiment. 
Define the Hilbert space
\[
\cH:=L^2(\R) \otimes \C^2 \,,
\] 
where the set of square integrable function encodes the momentum degree of freedom in transversal direction
and the two-dimensional component represents the spin.  We assume, for simplicity, that the only Hamiltonian that is relevant inside the furnace
is the Hamiltonian $H$ of the harmonic oscillator corresponding to the confining potential.
Hence,
the joint Hamiltonian of spin and momentum is then given by $H\otimes {\bf 1}$. 
In thermal equilibrium we have the state
\[
\rho:=\sum_n q_n |n\rangle \langle n |\otimes {\bf 1}\,,
\]
where $|n\rangle$ with $0=1,2,\dots$ denotes the eigenstates of the oscillator and $q_n$ the Boltzmann probabilities 
corresponding to the considered temperature. 

After the atoms leave the furnace, the
oscillator potential is no longer effective and  the system is no longer in equilibrium.
The inhomogeneous field generates a dynamics that entangles
spin and translational degree of freedom. We assume that the position degrees of freedom corresponding to other directions than
the transversal direction under consideration are irrelevant and we have free motion in longitudinal direction. 
 
When the atom arrives at the screen its state has been transformed
to $U (\rho\otimes {\bf 1}) U^\dagger$ with the unitary map
\[
U:=U_\downarrow\otimes |\downarrow \rangle \langle \downarrow |
+U_\uparrow \otimes |\uparrow \rangle \langle \uparrow |\,,
\]
where $|\downarrow\rangle,|\uparrow\rangle$ denote the two possible spin states and $U_\downarrow, U_\uparrow$
are unitary operators that act on the transversal degree of freedom in a spin-dependent way. 
When the atoms leave the furnace and enter the  field, a unitary dynamics transforms the state, i.e., the system is no longer in equilibrium. 
The relation between cause and effect in this example is therefore generated in a non-equilibrium dynamics, i.e.,
by removing the constraints like the  oscillator  potential.

To see how the form of the relevant quantum states is related  to the non-equilibrium
dynamics, we add the following observations. 
The states 
\[
U_\uparrow \rho U^\dagger_\uparrow  \quad \hbox{ and } \quad    U_\downarrow \rho U^\dagger_\downarrow
\]
are Gibbs equilibrium states for the transformed Hamiltonians
\[
U^\dagger_\uparrow H U_\uparrow \quad \hbox{ and } \quad U^\dagger_\downarrow H U_\downarrow\,.
\]
 If we assume that $U_\uparrow$ and $U_\downarrow$
are {\it simple} dynamical evolutions like translations, these are, again, simple Hamiltonians. 
Hence the conditional state of the system representing the effect, given a fixed value of the cause variable,
is a Gibbs state for a {\it simple} Hamiltonian. 

 On the other hand, the marginal state of the effect system itself is given by
$(U_\uparrow \rho U_\uparrow +U_\downarrow \rho U_\downarrow)/2$. 
The formal Hamiltonian that can be obtained from 
the logarithm of such a mixture, does not have any direct physical meaning\footnote{Jaynes stated in the context of
non-equilibrium thermodynamics \cite{JaynesGibbs}: ``[...] we must learn how to construct ensembles which describe not only the present values of macroscopic quantities, but also whatever information we have about their past behavior.''} and need not be simple.

If cause and effect are represented by the states of two physical systems 
(at the same time instant),
one influencing the other with negligible back action,
we are faced with the question whether such kind of causal unidirectionality already requires thermal non-equilibrium. 
To discuss this,
we revisit
the setting of Subsection~\ref{Arm} 
but with $T_1=T_2=T$. Then the joint
distribution of the systems reads
\[
p(x,y)=\frac{e^{-\beta H(x,y)}}{\sum_{x,y} e^{-\beta H(x,y)}}\,.
\]
Recall now the sender/receiver protocol  where system 1 was randomly adjusted to some value $x$ according to
the marginal distribution $p(x)$.  The conditional $p_\rightarrow (y|x)$ will then coincide with the usual equilibrium conditional $p(y|x)$. 
Hence the intervention preserves the usual equilibrium state. For symmetry reasons, this holds clearly for adjusting system 2, too. 
But then the backward and the forward information 
coincide exactly, i.e.,
\[
I_\rightarrow(X:Y)=I_\leftarrow(X:Y)\,.
\]
Hence, different temperatures were really needed in Subsection~\ref{Arm} to obtain a definite causal direction.

We want to revisit the second example in Section~\ref{PlMK} (with the 
spins  in a magnetic field)  in light of this result. The interaction between the field and the spin cannot be an interaction between
two systems in {\it Gibbs equilibrium with a common temperature}, otherwise the field would be influenced by the probe 
spin in the same way as vice versa,
in contradiction to our assumption on the definite causal direction. 
To show this, we assume that the field 
is generated by $n$ spin 1/2 particles.
Let
\[
S_z:=\sum_{j=1}^n \sigma_z^{(j)} \,.
\]
be their total spin in $z$ direction, where $\sigma_z^{(j)}$ denotes the Pauli matrix $\sigma_z$ on spin $j$.
The free Hamiltonian of the $n$-spin system
when subjected to a magnetic field $B$ in $z$ direction is given by
\[
H:= BS_z \otimes {\bf 1} \,.
\]
The free
Hamiltonian of the probe spin system  is $B({\bf 1} \otimes \sigma_z)$.

In its thermal  equilibrium, the total spin $S_z$
follows a binomial distribution
$B_q(k)$
with $q/(1-q)=\exp(-1/kT)$.
For large $n$,  the  total magnetic moment
fluctuates on the scale $\sqrt{n}$.
Then we introduce an interaction $H_i$ by
\[
H_i:= c \frac{1}{\sqrt{n}} S_z \otimes \sigma_z\,,
\]
with a constant $c$ determining the interaction strength.
The scaling
factor $1/\sqrt{n}$ is chosen such that the total field strength
``felt'' by the probe spin system
follows  a well-defined distribution in the limit $n\to\infty$.
Now we consider the conditional probability for $k$ spins up given that the
probe spin is in its upper state. The total spin of the $n$-particle system defines an integer-valued random variable $Y$.
We have
\[
p(y=k|x=1/2)=B_{q^+}(k)
\]
and
\[
p(y=k|x=-1/2)=B_{q^-}(k)
\]
with $q^\pm$ are given by $q^\pm/(1-q^\pm)=\exp((1\pm \sqrt{1/n})/kT)$.
In the limit of large $n$ the binomial distributions can be replaced with
two Gaussians with standard deviation in the order of $\sqrt{n}$.
Their mean value differ also on the scale $\sqrt{n}$. This shows that
we obtain a mixture of two Gaussians for the distribution of magnetic moments
of the $n$-particle system. Once we adjust the probe spin, bimodality disappears.
This shows that we have mutual influence  between probe and the system generating the field. 

We conclude that
every example discussed in this paper relies on non-equilibrium states.

\section{Conclusions}

We have described several physical settings where the conditional
probability for an effect given its cause is less complex than the
probability for the cause given its effect. 
Here we have considered
different notions of complexity, e.g., hierarchy of exponential families
as well as with respect to computational complexity.

To link this kind of ``asymmetric Occam's Razor principles'' with the thermodynamic arrow of time
we have constructed models where the statistical asymmetries between cause and effect are implications of the
irreversibility of mixing processes. The common root between all the known asymmetries is therefore the 
tendency of {\it specific} initial conditions to evolve into {\it typical} final states.
Specific initial conditions can, for instance, be product probability distributions of joint systems that evolve typically to
distributions with stochastic dependences. The fact that specific initial conditions occur more often than 
specific final conditions is linked with the second law, or may even be considered as its essential content.

However, appropriate notions of simplicity 
have yet to be discovered. Since it is impossible to draw reliable causal conclusions
from statistical observations that do not involve interventions, we have to restrict ourselves to finding 
causal inference rules which are {\it often} valid. These have to be based upon observing which transition probabilities
$P({\rm effect}|{\rm cause})$ are likely to occur in nature and which ones are likely to correspond to non-causal conditionals. 
To explore this asymmetry in a systematic way 
as well as its   relation to the thermodynamics of irreversible processes
is an important challenge for both machine learning and theoretical physics.

\section*{Acknowledgements}

The author would like to thank Armen Allahverdyan  for helpful discussions and comments, which were especially essential for
Subsection~\ref{Arm}.

\end{document}